\documentclass[12pt]{article}
\usepackage[margin=0.9in]{geometry}
\usepackage{amsmath,amsthm,amssymb}
\usepackage{setspace}
\usepackage{bm}
\usepackage{mathrsfs}
\usepackage{comment}
\usepackage{graphicx}
\usepackage{booktabs}
\usepackage[table]{xcolor}
\usepackage{xcolor} 
\usepackage{hyperref}
\usepackage{multirow, multicol}
\usepackage{natbib}
\setcitestyle{authoryear,open={[},close={]}}

\newtheorem{proposition}{Proposition}

\theoremstyle{definition}
\newtheorem{remark}{Remark}

\hypersetup{
    colorlinks=true,            
    linkcolor=teal,             
    citecolor=blue!70,    
    filecolor=magenta,          
    urlcolor=cyan               
}

 \newcommand{\RR}{\mathbb{R}}
 \newcommand{\YY}{\mathcal{Y}}
 
 \newcommand{\iid}{\stackrel{\mathrm{iid}}{\sim}}
 \newcommand{\Ga}{\text{Gamma}}
  \newcommand{\Normal}{\text{Normal}}
\newcommand{\unif}{\text{Uniform}}
\newcommand{\DP}{\text{DP}}
\newcommand{\CRM}{\text{CRM}}

\newcommand{\SD}{\text{sd}}
\renewcommand{\th}{\theta}

\newcommand{\bx}{\bm{x}}

\newcommand{\tmu}{{{\mu}}}
\newcommand{\tmus}{{{\mu}^\star}}
\newcommand{\tht}{\tilde{\th}}
\newcommand{\zt}{\widetilde{z}}
\newcommand{\Gx}{G_x}
\newcommand{\Jt}{\widetilde{J}}
\newcommand{\scf}{\mathcal{F}}
\newcommand{\sx}{\mathcal{X}}
\newcommand{\sy}{\mathcal{Y}}
\newcommand{\bth}{\bm{\theta}}
\newcommand{\E}{\mathbb{E}}
\newcommand{\R}{\mathbb{R}}
\renewcommand{\scf}{\mathcal{F}}
\renewcommand{\sx}{\mathcal{X}}
\renewcommand{\sy}{\mathcal{Y}}

\newcommand{\sd}{\mathcal{D}}
\newcommand{\eps}{\epsilon}

\newcommand{\ub}{\mathbf{u}}
\newcommand{\zb}{\mathbf{z}}

\newcommand{\zs}{z^\star}
\newcommand{\ns}{n^\star}

\newcommand{\thb}{\bm{\th}}
\newcommand{\muc}{\tmu^o}
\newcommand{\nuc}{\nu^o}

\newcommand{\schwarz}{\color{black}}

\newcommand{\ech}{\schwarz\rm}

\newcommand{\mypar}[1]{}

\newcommand{\red}{\color{red}}

\newcommand{\bla}{\color{black}}
\newcommand{\rest}{\ldots}
\newcommand{\sumni}{\sum_{i=1}^n}

\newcommand{\frall}{\text{ for all }}

\newcommand{\XX}{200 }

\title{\bf \LARGE DPGLM: A Semiparametric Bayesian  GLM \\ with Inhomogeneous
  Normalized\\ Random Measures}

\author{
    \begin{tabular}{c}
        \bf \large Entejar Alam$^{1}$,  Paul J. Rathouz$^{2,1}$, and Peter M{\"u}ller$^{1,3}$ \\
        \normalsize  $^{1}${Department of Statistics and Data Sciences,} \\
       \normalsize $^{2}${Department of Population Health,} \\
        \normalsize $^{3}${Department of Mathematics,} \\
         \normalsize University of Texas at Austin, TX, USA
    \end{tabular}
}

\date{}

\begin{document}

\maketitle

\begin{abstract}
We introduce a novel varying-weight dependent Dirichlet process (DDP) model that extends a recently developed semi-parametric generalized linear model (SPGLM) by adding a nonparametric Bayesian prior on the baseline distribution of the GLM. We show that the resulting model takes the form of an \emph{inhomogeneous} completely random measure that arises from exponential tilting of a normalized completely random measure. Building on familiar posterior sampling methods for mixtures with respect to normalized random measures, we introduce posterior simulation in the resulting model. We validate the proposed methodology through extensive simulation studies and illustrate its application using data from a speech intelligibility study.
\vspace{1em}

\end{abstract}

\doublespacing

\section{Introduction}
We introduce a non-parametric Bayesian extension of the
semi-parametric GLM defined in \citet{rathouz2009generalized}. Under the proposed model, the
marginal distribution of the response, conditional on a given
covariate takes the (approximate -- to be made more precise
later) form of an inhomogeneous normalized random measure  (NRM)
\citep{regazzini2003distributional}. The joint model (across
covariates $x$) is a variation of the popular
dependent Dirichlet process (DDP) model 
\citep{maceachern;2000, quintana2022dependent}, replacing the
marginal DP by an exponentially tilted DP with varying weights across
covariates. We discuss the model construction, including representations as NRM and DDP models, and 
characterize the posterior law. Appropriate extensions of the results
in \citet{james2009posterior} allow for straightforward posterior
simulation. We validate the proposed model with a simulation study and
illustrate it with an application on speech intelligibility development in children across ages 30 to 96 months.  

We build on the semi-parametric GLM introduced in
\citet{rathouz2009generalized}. Consider a GLM
\begin{equation}
  p_x(y) \equiv p(y \mid x) \propto \exp (\th_x y)\tmu(y) \label{spglm:eq1} 
\end{equation}
with continuous response $y \in \YY \subset \RR$ and a
p-dimensional covariate vector $x \in \mathcal{X}$ and (log)
normalization constant 
\begin{eqnarray}
\label{eq:2}
b(\th_x) = \log \int_{\YY} \exp (\th_x y) \tmu(dy). \label{spglm:eq2}   
\end{eqnarray}
In anticipation of the upcoming discussion, we allow $\tmu(y)$ to
be an un-normalized positive measure, implying
a  baseline density (i.e., when $\th_x=0$ in \eqref{eq:2}) $f_{\tmu} = \tmu/\tmu(\YY)$
in the GLM \eqref{spglm:eq1}. While in the classical GLM, the baseline
distribution is assumed to be in a parametric family,
in the semi-parametric SPGLM model the measure $\tmu(y)$
itself becomes an unknown parameter. As in the classical GLM, we
introduce a linear predictor $\eta = x^T\beta$, and a link function
$g$ to implicitly define $\th$ by requiring $\lambda \equiv \E(y \mid
x) = g^{-1}(\eta)$. 
That is,  
\begin{equation}
\lambda(x) = \E(y \mid x) = b^{\prime}(\th_x) = \int_{\YY} y \exp \{\th_x y - b(\th_x)\} \tmu(dy). \label{spglm:eq3}
\end{equation}
Noting that, for given $\tmu$, $b^{\prime}(\theta)$  is a strictly increasing
function of $\theta$ and invertible we have
$\th_x = {b^\prime}^{-1}(\lambda; \tmu)\stackrel{def}{=}\theta(\beta, \tmu, x)$. Here
we added $\tmu$ to the arguments of ${b^\prime}^{-1}$ to highlight the
dependence on $\tmu$. Alternatively, when we want to highlight
dependence on $\beta$ and $x$, indirectly through $\lambda$, we write
$\th_x = \th(\beta, \tmu, x)$. 

The defining characteristic of the SPGLM is a nonparametric baseline
or reference distribution $f_{\tmu}$ that replaces a parametric
specification in the classical GLM such as a binomial or Poisson
model. Keeping $f_{\tmu}$ nonparametric instead, the analyst
needs to only specify the linear predictor and link function, even
avoiding a variance function, leaving model specification less onerous than
even with quasi-likelihood (QL) models, while still yielding a valid
likelihood function.
Beyond the initial introduction of the SPGLM by
\citet{rathouz2009generalized}, which focused primarily on the
finite support case,  
\citet{huang2014joint} characterized the SPGLM in the infinite support case, and
\citet{maronge2023generalized} discussed the use with outcome-dependent or generalized case-control sampling. For a finite support case, \citet{alam2025dir} introduced a parametric Bayesian approach by assuming a Dirichlet distribution prior on $f_\mu$.
In \citet{rathouz2009generalized} and \citet{wurm2018semiparametric}, the SPGLM model is referred to as generalized linear density ratio model (GLDRM), and \citet{wurm2018semiparametric} fully develop the current ML computational algorithm and, including a working package \texttt{gldrm} on CRAN \citep{gldrm_package}.
Despite these developments, there are still many important gaps in the
literature. These include inference for application-driven functionals of
the fitted models such as exceedance probabilities, which are crucial
in clinical diagnosis \citep{paul2021dynamics}; natural hazard
detection \citep{kossin2020global}; financial risk management
\citep{taylor2016using}; conditional quartiles \citep{davino2022composite} or in general, any decision-making
setting. These inference problems are not straightforward to address with
maximum likelihood based approaches. In this paper, we set the table to address these gaps by developing a non-parametric Bayesian (BNP)
extension of the SPGLM.
In this BNP model we introduce $\tmu$ as an (un-normalized) positive
random measure. We use a prior on $\tmu$ to implicitly define an exponentially
tilted DP prior for $p_x$ in \eqref{spglm:eq1}.

In Section \ref{subsec:DPGLM}, we introduce the proposed semiparametric Bayesian extension of the
SPGLM, referred to as DPGLM, building on the finite version of \citet{alam2025dir}. We characterize the proposed model as a variation of the popular DDP model in Section \ref{subsec:DDP}, and in Section \ref{subsec:NRM} we show a representation of the implied marginal for one covariate as an inhomogeneous NRM.
In Section \ref{sec:post} we characterize the posterior distribution
under the DPGLM by showing it to be conditionally conjugate given
auxiliary variables similar to the construction used in
\citet{james2009posterior}.  
Section \ref{sec:sim} summarizes a simulation study. Section \ref{sec:ex} discusses an application, and Section \ref{sec:disc} concludes with a final discussion.

\section{The DPGLM Model} 
\label{sec:DPGLM}

\subsection{A Bayesian semiparametric GLM}
\label{subsec:DPGLM}
We extend
\eqref{spglm:eq1}--\eqref{spglm:eq3} to a Bayesian inference model by
adding a prior probability model for all unknown parameters, including in
particular the baseline density $f_{\tmu}(\cdot) \equiv \tmu \big/\tmu(\sy)$.
Prior models for random probability measures like $f_{\tmu}$
are known as non-parametric Bayesian models (BNP)
\citep{ghosal2017fundamentals}. 
The most widely
used BNP model is the Dirichlet process (DP) prior introduced in the
the seminal work of \citet{ferguson1973bayesian}. 
The DP prior is characterized by two parameters: a
concentration parameter $\alpha$ and a base distribution $G_0$.
We write $G \sim \DP(\alpha, G_0)$.
One of the many defining properties of the DP is the 
stick-breaking representation of 
\citet{sethuraman1994constructive} for $G \sim \DP(\alpha, G_0)$ as 
\begin{equation}
G(\cdot) \equiv \sum_{h=1}^\infty s_h \delta_{\zt_h}(\cdot)
\label{eq:stickb}
\end{equation}
with atoms $\zt_h \iid G_0$, and weights $s_h = v_h
\prod_{\ell<h}\left(1-v_{\ell}\right)$, where $v_h \iid Be(1,
\alpha)$.

An alternative  defining property of the DP prior is as a normalized
completely random measure. A completely random measure (CRM) is a
random measure $\tmu$ with the property that the random measures assigned to any two non-overlapping events $A,B$ are independent, that is $\tmu(A) \perp
\tmu(B)$ when $A \cap B = \emptyset$
\citep{kingman1967completely}. 
A CRM is characterized by its Laplace transform $\E[\exp\{-\int h(y)
\tmu(dy) \}]$ for any measurable function $h$, which in turn is
completely characterized by the L\'evy intensity $\nu(ds, dy)$ that
appears in the L\'evy-Khintchine representation 
\begin{equation}
\E\left\{e^{-\int_\sy h(y) \mu(dy)}\right\} = \exp \left[- \int_{R^+
    \times \sy} \left\{1 - e^{-sh(y)}\right\}\nu(ds, dy) \right].
\label{eq:LK}
\end{equation}
If $\nu$ factors as $\nu(s,y)= \rho(s)\, G_0(y)$ the CRM is known as
a homogeneous CRM. \citet{regazzini2003distributional} introduced the
wide class of normalized random measures (NRM) by defining a BNP prior
for a random probability measure $f_{\tmu}$ as
${\tmu}/{\tmu(\sy)}$, with  
a CRM $\tmu$.
The DP is a special case of NRM, using a normalized gamma CRM with L\'evy intensity
\begin{equation}
  \nu(ds, dy)= \frac{e^{-s}}{s} ds \cdot \alpha G_0(dy)
\label{eq:ga}
\end{equation}
for a $DP(\alpha, G_0)$. We use a gamma CRM as prior model for $\tmu$ in
the SPGLM \eqref{spglm:eq1}, with base measure $G_0$ on the 
support $\YY$ and concentration parameter $\alpha$,
implying a DP prior on the baseline density $f_{\tmu}$. We add a normal
prior on $\beta$ to complete the prior specification
\begin{align} 
\label{priors}
  &\tmu \sim \text{Gamma CRM}(\nu)
    \mbox{ with }
    \nu(ds, dy) = \frac{e^{-s}}{s} ds \cdot \alpha G_0(dy) \nonumber \\ 
    & \beta \sim \text{MVN}(\mu_\beta, \Sigma_\beta) .
\end{align}
The two lines of~\eqref{priors} jointly imply a prior on $\scf = \{p_x: x \in \sx\}$. 
We add one more extension by adding a convolution with a continuous kernel $K(y_i \mid
z_i)$ and a latent variable $z_i$ to define a continuous sampling model for $y$.
Using a symmetric kernel $K(\cdot)$, this does not change the mean
regression structure of the GLM, as
$\E(y_i \mid x_i) =
\E_{z_i \mid x_i}\left\{  \E(y_i \mid x_i, z_i) \right\} =
g^{-1}(x^\prime_i \beta)$. 
For reference, we state the complete hierarchical model: In the model statement we introduce notation $\Gx(z)$ for the sampling model
 $p(z \mid x)$ for the latent $z_i$ (similar to $p_x$ for observed
 $y_i$ in \eqref{spglm:eq1}).
\begin{align}
\label{DPGLM}
  y_i \mid  z_i & \sim K(y_i \mid z_i),
                  \text{ conditionally independent of } x_i,  \tmu, \beta\\
  z_i \mid x_i = x,  \tmu,\beta & \sim \Gx(z_i) \propto
                                       \exp(\th_{x} z_i) \tmu(z_i), 
                  \text{ with } {b^\prime}(\th_x) =
                  g^{-1}(x^\prime \beta) \nonumber \\  
  \tmu & \sim \text{Gamma CRM}(\nu), \text{ with } \nu(ds, dz) =
         \frac{e^{-s}}{s} ds \cdot \alpha G_0(dz) \nonumber \\  
  \beta & \sim \text{MVN}(\mu_\beta, \Sigma_\beta)\ .  \nonumber
\end{align}
Recall that $\th_x=\th_x(\beta,\tmu)$ is a derived parameter.
We refer to the proposed model \eqref{DPGLM} as 
DPGLM. 
Also, we refer to  
$\tmu_i = \tmu(z; \theta_{x_i}) \equiv \exp(\th_{x_i} z) \tmu(z)$ as the \emph{tilted CRM}, with tilting parameter
$\theta_{x_i}$.  

Finally, a note on identifiability in model \eqref{DPGLM}.
Consider a pair $\tmu, \{\th_x; \; x \in X\}$, and another one with
$\tmu' \equiv \tmu\cdot e^{cz}$ and $\{\th_x'=\th_x-c\}$.
All else being equal, the two sets of parameters have identical likelihood.
For a meaningful report of inference on $\tmu$ we will use
post-processing to replace $\tmu$ with $\tmu \equiv \tmu \cdot
e^{cz}$, with $c$ to ensure $\int z\, df_{\tmu}(z) = m_0$ for a fixed $m_0$, specified by the analyst.
An interesting alternative could be to restrict the prior on $\tmu$
using a generalized notion of conditioning a DP prior that is
introduced in current work by \citet{lee2024}.

\subsection{A varying weights DDP}
\label{subsec:DDP}

\citet{maceachern;2000} first introduced the
\textit{dependent} Dirichlet process (DDP) by extending the 
DP model to a family of random distributions $\{G_x: x \in \sx \}$.
The construction starts by assuming marginally, for each $x$, a DP
prior for each $G_x = \sum w_{xh} \delta_{m_{xh}}$.
The desired dependence can then be accomplished by using shared
$w_{xh}=w_h$ and defining a dependent prior for $\{m_{xh},\; x \in
X\}$ while maintaining independence across $h$, as required for the
marginal DP prior. This defines the {\it common weights DDP}.
Alternatively one can use common atoms $\zt_h$ with a dependent prior on
varying weights $\{w_{xh},\; x \in X\}$ ({\it common atoms DDP}), or
use varying weights and atoms. See, for example,
\citet{quintana2022dependent} for a review of the many different
instances of DDP models.
A commonly used version are common weights and Gaussian process
(GP) priors for $\{m_{xh},\; x \in X\}$, independently across $h$
\citep{maceachern;2000}.

In the proposed DPGLM approach (\ref{DPGLM}), dependence is
introduced naturally through weights $w_{xh}$ (defined below) while keeping atoms $\zt_h$ constant across $x$.
Starting from the representation
\eqref{eq:stickb}  for a (single) DP prior we define
$\Gx(z)$ as follows:  
\begin{multline}
\Gx(z)  = \exp\left\{\th_x z - b(\th_x)\right\} \tmu(z) 
=
\exp\left\{\th_x z - b(\th_x)\right\} \sum_{h=1}^\infty s_h
      \delta_{\zt_h} (z)   \\
 = 
\sum_{h=1}^\infty \left[\exp\left\{\th_x \zt_h - b(\th_x)\right\}
      s_h\right] \delta_{\zt_h} (z) 
 =  
      \sum_{h=1}^\infty  w_{xh} \delta_{\zt_h} (z),
      \label{eq:pxz}
\end{multline}
where $w_{xh} = \exp\left\{\th_x z - b(\th_x)\right\}  s_h$,
depends on $x$ implicitly through $\th_x$. That is, $w_{xh}$ are introduced by exponential tilting of one
random measure $\tmu$ which is shared across all $x$. 
The model defines a
variation of a DDP model using common atoms and varying
weights. However, the exponential tilting in \eqref{eq:pxz} defines a marginal prior $\Gx$ beyond a DP model, as we shall 
discuss next in more detail.

\subsection{The marginal model}
\label{subsec:NRM}
The implied marginal model $\Gx(z)$ for given covariate $x$ in
\eqref{DPGLM} can be shown to be an NRM again.
This is seen by noting that the Laplace transform of $\Gx$ takes the
form of \eqref{eq:LK} again, allowing us to recognize the NRM by
inspection of the L\'evy intensity in \eqref{DPGLM}. 
\begin{proposition}
     \label{result1}
  \citep{nieto2004normalized}
  Consider the DPGLM with implied marginal distribution $\Gx(z)
  \propto \exp(\th_{x} 
  z) \tmu(z)$, assuming a gamma CRM \eqref{eq:ga}, i.e.,
  $\frac{\tmu}{\tmu(\YY)} = f_{\tmu} \sim \DP(\alpha, G_0)$, and given $\th_x$. Then $\Gx$ is an inhomogeneous normalized random measure (NRM) with L\'evy intensity,
  \begin{equation}
  \nu(ds, dz) = \frac1s\,{e^{- s\, \exp (-\th_x z)}} ds \cdot \alpha
  G_0(dz)
  \label{eq:res1}
  \end{equation}
\end{proposition}
The L\'evy intensity $\nu$ in \eqref{eq:res1} characterizes an
inhomogeneous NRM, with $\rho(ds \mid z) = \frac1s\, {e^{- s\, \exp(-\th_x z)}}\; ds$ varying with $z$.  

The use of the DP prior for $f_{\tmu}$ makes the result in
\eqref{eq:res1} particularly simple, allowing a closed form
expression. A similar result, albeit not necessarily in closed form
anymore, is true under any other NRM prior for 
$f_{\tmu}$. For example, \citet{lijoi&mena&pruenster:07} argue for the
richer class of normalized generalized gamma, which includes the DP as
a special case.
One common reason to consider alternatives to the DP prior is the lack
of flexibility in modeling the random partition implied by ties of a
sample from a DP random measure. In more detail, in the context of \eqref{DPGLM} the
discrete nature of $\tmu(\cdot)$ gives rise to ties among the
$z_i$. Using a DP prior, under $\th_x=0$ the random partition characterized by the
configuration of ties is known as the Chinese restaurant process. It
is indexed by a single hyperparameter, $\alpha$.
\citet{de2013gibbs}, for example, argue that the nature of this random
partition is too restrictive for many applications.
However, in the context of the DPGLM, the random partition is not an
inference target, and we shall never interpret the corresponding
clusters, leaving the DP prior as an analytically and computationally
appealing prior choice for $\tmu$.

The BNP prior for $\Gx(z)$ and the kernel in the first two levels of
the DPGLM model \eqref{DPGLM} define a variation of popular BNP
mixture models. The use of the particular NRM with L\'evy intensity
\eqref{eq:res1} arises naturally in the context of the GLM-style
regression with the exponential tilting.
Posterior simulation for BNP mixtures with NRM priors on the mixing
measure is discussed, for example, in
\citet{Argiento:2010}, \citet{barrios&al:13} or \citet{favaro&teh:13}.
However, the GLM regression introduces a complication by applying
different exponential tilting for each unique covariate $x_i$.
This leads to some variations in the posterior characterization and
the corresponding posterior simulation algorithms. We next discuss those changes. 

\section{Posterior characterization}
\label{sec:post}
Let $\sd_n = \{x_i, y_i\}_{i=1}^n$ denote the observed data, where $x_i
\in \sx \subset R^p$ and $y_i \in \sy \subset R$,
and (\ref{DPGLM}) adds the latent variables  $z_i$. 
For simplicity we write $\theta_i$ for $\theta_{x_i}$ and define
$T_i = \int_{\YY} \exp (\th_i z) \tmu(dz)$ as the total mass of the 
tilted, un-normalized CRM $\tmu_i = \tmu(z; \theta_i) \equiv \exp(\th_i z) \tmu(z)$. We then adapt the results in Section 2 of
\citet{james2009posterior}
to characterize the posterior distribution under the DPGLM model
(\ref{DPGLM}).

We first introduce a data augmentation scheme with auxiliary variables $u_i$, using one auxiliary variable for each unique covariate vector $x_i$. For the moment, we assume that all $n$ covariate vectors are distinct (and shall comment later on simple modifications to accommodate the more general case). We define
$$
u_i \mid T_i \equiv T(\theta_i, \tmu) \sim \gamma_i/T_i,
$$
where $\gamma_i \sim \operatorname{Exp}(1)$ are independent across $i$ and also from $T_i$, implying $p(u_i \mid T_i) = \Ga(1, T_i)$. Recall that as a normalizing constant, $T_i$ is a function of all model parameters including $\tmu$. We next state the posterior for $\ub = (u_1, \dots, u_n)$, conditional on the latent variables $\zb = (z_1, \dots,
z_n)$ but marginalizing w.r.t. $\tmu$ (and thus $T_i$), and for fixed $\bth = (\th_1, \dots, \th_n)$ (by a slight abuse of notation, we include $\thb$ in the conditioning subset to emphasize that it is held fixed). 
\begin{proposition}
\label{result2}
Let $\bth = (\theta_1, \dots, \theta_n)$ and $\zb = (z_1, \dots,
z_n)$. Then
$$
p(\ub \mid \bth, \zb) \propto \exp \left \{ - \int_{\sy} \log \left(1+ \sum_{i=1}^n u_i e^{\th_i  v} \right) G_n(dv) \right\}, 
$$
where $G_n = \alpha G_0 +  \sum_{i=1}^n \delta_{z_i}$.  
\end{proposition}
The proof of Proposition \ref{result2} is implied as part of the proof for the next Proposition. Complete proofs for Propositions~\ref{result1}--\ref{result3} are provided in Appendix \ref{sec:proofs}.  As mentioned previously, the discrete nature of $\tmu$ introduces ties in $z_i$. Let $\{z^\star_1, \dots, z^\star_k\}$ denote the unique values, with multiplicity $\{n^\star_1, \dots, n^\star_k\}$, among the currently imputed $\{z_1,\ldots,z_n\}$. Clearly, $\sum_{\ell=1}^k n^\star_\ell = n$. Then $G_n$ in Proposition \ref{result2} can be written as $G_n = \alpha G_0 + \sum_{\ell = 1}^k n^\star_\ell \delta_{z^\star_\ell}$.

Conditional on $\ub$ and for fixed $\thb$, the posterior distribution of $\tmu$ is an inhomogeneous CRM. In particular, we have:
\begin{proposition}
\label{result3}
Let $\bth = (\theta_1, \dots, \theta_n)$ and $\zb = (z_1, \dots,
z_n)$ with $k$ unique values $z^\star_1, \dots, z^\star_k$ having multiplicities $n^\star_1, \dots, n^\star_k$. Then $\tmu$ includes atoms at the $\zs_\ell$ with random probability masses $J_\ell, \, \ell = 1, \dots, k$. Letting $\muc$ denote the remaining part of $\tmu$ we have  
$$
\tmu \mid \ub, \zb, \bth \stackrel{d}{=}
\muc + \sum_{\ell=1}^k J_\ell \delta_{z^\star_\ell}, 
$$
where:
\begin{enumerate}
\item[1.]  $\muc \sim \CRM\left(\nuc\right)$
   with the L\'evy intensity 
   $$\nuc(ds, dz) = \frac1s\,
  {\exp\left\{-s\left(1 + \sum_{i=1}^n u_i e^{\th_i z}\right)\right\}} ds \, \alpha G_0(dz).$$ 
 \item[2.] Let
   $\psi(z^\star_\ell; \, \ub, \bth) = \sum_{i=1}^n u_i
   e^{\th_i z^\star_\ell}$. Then, for $\ell = 1, \dots, k$,
   $$
   P_{J_\ell}\left(s \mid \ub,  \bth, z^\star_\ell, \ns_\ell\right) 
   \propto s^{\ns_\ell - 1}\, e^{-s \{1+\psi(z^\star_\ell; \, \ub, \bth)\}}
   \equiv \Ga\left(\ns_\ell, \, \psi(z^\star_\ell; \ub, \bth) + 1\right).
   $$
 \end{enumerate}
 and $\muc$ and $J_\ell$ are independent given $\ub, \zb$. 
\end{proposition}
Proposition \ref{result3} shows that, given $\zb$ and $\ub$ (with fixed $\bth$), {\em a posteriori} $\tmu$ is again a CRM. To be precise, it is a sum of two components. One part is an inhomogeneous CRM $\muc = \sum_{\ell = 1}^\infty \Jt_\ell \delta_{\zt_\ell}$ with L\'evy intensity $\nuc$.
The random atoms $\zt_\ell$ and weights $\Jt_\ell$ can be generated using, for example, the \citet{ferguson1972representation} algorithm.
The second component is a finite discrete measure with gamma distributed random weights $J_\ell$ at fixed atoms $\zs_\ell$. The latent variables $z_i$ are updated via their complete conditional distribution: $p(z_i\mid \tmu,\theta_i)\propto K(y_i\mid z_i)\sum_{\ell}\exp(\theta_i z_i)\bar{J}_\ell \delta_{\bar{z}_\ell}(z_i)$, where $\{\bar{z}_\ell\}_{\ell \geq 1}=\{\tilde{z}_\ell\}_{\ell \geq 1}\cup \{z^\star_\ell\}_{\ell=1}^k$ and $\{\bar{J}_\ell\}_{\ell \geq 1}=\{\tilde{J}_\ell\}_{\ell \geq 1}\cup \{J_\ell\}_{\ell=1}^k$. 

There is one important detail about Proposition \ref{result3} and \ref{result2}. The result holds for fixed $\bth$.
But in \eqref{DPGLM} we use instead the derived
parameter $\th_x = \th_x(\tmu)$. This adds 
more information on $\tmu$, indirectly through $\th_x(\tmu)$. 
However, the result of Proposition~\ref{result3}
hinges on independent sampling with given, fixed exponential tilting, and is not easily extended to using $\th_x(\tmu)$. 
Instead, we exploit Proposition~\ref{result3} to implement a Metropolis-Hastings (MH) transition probability.  
Let $\tht_x=\th_x(\tmu)$ denote the derived parameters implied by the currently imputed CRM $\tmu$. Let then $q(\tmus \mid \tmu)$ denote the inhomogeneous CRM described in Proposition \ref{result3} with fixed $\th_x=\tht_x$. That is, the described distribution on $\tmus$ under fixed $\bth$ implied by the currently imputed $\tmu$. We then treat $\tmus$ as a proposal in a MH transition probability and follow up with an MH step with acceptance ratio $r$ (details for evaluating $r$ are provided in Appendix \ref{sec:mcmc}).

Finally, in the general case with possible ties of the covariate vectors $x_i$, one could still use the same results, with $n$ auxiliary variables $u_i$. Alternatively, the following construction could be used with fewer auxiliary variables. Let $x^\star_j$, $j=1,\ldots,J$ denote the unique covariate combinations with multiplicities $a_j$. Let then $T_j$ denote the normalization constant under covariate $x=x^\star_j$. Similar results as above hold, starting with latent variables $u_j \sim \Ga(a_j, T_j), \, j = 1, \dots, J$.

We list details of the posterior MCMC simulation in Appendix
\ref{sec:mcmc}. Finally, for reporting posterior inference, we use post-processing to address the lack of likelihood identifiability for $f_{\tmu}$ (recall the note at the end of Section \ref{subsec:DPGLM}). Specifically, to obtain more meaningful posterior summaries, we report inference on  $f_{\tmu}$ subject to the constraint $\int y\, df_{\tmu}(y) = m_0$ for a fixed $m_0$; that is, we tilt the posterior samples of $f_\mu$ so that they all share a common mean $m_0$. In practice, the choice of $m_0$ can be based on prior judgement, or alternatively, one can use any measure of central tendency of $y$, such as the mean or median.

\section{Simulation Studies}
\label{sec:sim}
We proceed with simulation studies to evaluate the frequentist inference operating characteristics under the DPGLM model. We aim to address the following key questions: (Q1) How accurately does the model estimate the baseline density, $f_{\tmu}(y)$, the baseline cumulative distribution function, $F_{\tmu}(y)$, and exceedance probabilities, $p(y > y_0 \mid x)$, under various scenarios? (Q2) Do the 95\% credible intervals for $f_{\tmu}(y)$, $F_{\tmu}(y)$ and exceedance probabilities achieve nominal coverage rates?
(Q3) Do the credible intervals for the regression parameters $\beta_j$ attain nominal coverage? How does the predictive performance of the DPGLM compare with that of a competing Beta regression model?

We study these questions under realistic sample sizes and
simulation truths mimicking the data analysis presented later, in Section \ref{sec:ex}, using the Speech Intelligibility dataset, with continuous outcomes $y \in \sy \subset (0,1)$.

\paragraph*{Data generating mechanism.} 
For each setting, we generate covariates $x_i \sim \unif(a, b)$, with $a = {-\sqrt{12}}/{4}$ and $b= {\sqrt{12}}/{4}$. This choice ensures that $\SD(x_i) = 1/2$. We use $\mathcal{D}_n$ to refer to the observed data $\{x_i, y_i\}_{i=1}^n$. We consider the following two scenarios: 
\begin{itemize}
    \item \emph{\textbf{Null case (scenario I):}} The baseline density $f_\tmu$ is set equal to a kernel density estimate, $f^{(kde)}_{\tmu}$, computed from the Speech Intelligibility dataset (ignoring covariates). Then, in this scenario, the responses $y_i$ are sampled independently of $x_i$; that is, no covariate effect is present. Consequently, the regression parameters are effectively $\beta_0 = 1$ (intercept) and $\beta_1 = 0$ (slope). Specifically, we sample $y_i \sim p(y_i) \propto \exp(\theta y_i)f^{(kde)}_{\tmu}(y_i)$, where $\th = {b^\prime}^{-1}\left\{g^{-1}(\beta_0)\right\}$. Here, the link function $g$ is taken to be the \emph{logit} function, and $b^\prime$ is defined as in (\ref{spglm:eq3}).

    \item \emph{\textbf{Regression (scenario II):}} In this scenario, the baseline density remains the same as in the foregoing scenario, $f^{(kde)}_{\tmu}$, computed from the Speech Intelligibility dataset, but the responses now depend on the covariates. Specifically, we sample $y_i \sim p(y_i \mid x_i) \propto \exp(\theta_{x_i} y_i)f^{(kde)}_{\tmu}(y_i)$, where the covariate effect is incorporated through $\th_{x_i} = 
      {b^\prime}^{-1}\left\{g^{-1}(\eta_{x_i})\right\}$, with $\eta_{x_i} = \beta_0 + \beta_1 x_i$. Again, the link function $g$ is the \emph{logit} function, and $b^\prime$ is defined as in (\ref{spglm:eq3}). We set $\beta_0 = 0.2$ and $\beta_1 = 0.7$. 
\end{itemize}
In the following discussion, recall that under the DPGLM, after the additional convolution described in (\ref{DPGLM}), the baseline density is given by, $f_{\tmu}(y) = \int_\sy K(y \mid z) \mu(dz) / \mu(\YY)$. For each setting and for sample sizes $n = 50, 100, 250$, we generate \XX datasets. For each data set, we fit the proposed DPGLM model using
a $\unif(z-c, z+c)$ kernel $K$, where $c$ is chosen proportional to Silverman's rule of thumb \citep{silverman1986}. The prior
distributions are specified as in (\ref{DPGLM}), with $\alpha = 1$ and $G_0$ = $\unif(0,1)$. We implement
MCMC posterior simulation for $\beta$ and $\tmu$ using the transition probabilities detailed in Appendix \ref{sec:mcmc}; a total of $2,000$ MCMC samples are generated for each data replicate, with the first 1,000 iterations discarded as burn-in and the remaining $R = 250$ samples, after thinning by a factor of 4, used for inference.

To formalize responses to questions Q1--Q3, we
employ the following performance metrics. For (Q1), we assess the goodness‐of‐fit in estimating $F_{\tmu}(y)$ by computing, for each data replicate, the \emph{Kolmogorov-Smirnov} (KS) test statistic, $D = \sup_{y \in \sy} |\widehat F_{\tmu}(y) -
F_{\tmu}(y)|$, and by calculating the \emph{total variation} (TV) distance for estimating $f_{\tmu}(y)$, defined as: $\text{TV} = \frac{1}{2}\int_\sy |\widehat f_{\tmu}(y) - f_{\tmu}(y)| \, dy$. Moreover, we compute the 1-\emph{Wasserstein} distance as \[
W_1 = \int_{\sy} \Bigl|\widehat F^{-1}_{\tmu}(y) - F^{-1}_{\tmu}(y)\Bigr|\,dy,
\] which compares the quantile functions of the estimated and true $F_{\tmu}(y)$. To ensure valid comparisons, we tilt each of the posterior samples of $f_{\tmu}(y)$ to have a common reference mean $m_0$, set equal to the simulation ground truth computed from $f^{(kde)}_{\tmu}$. For (Q2), we first compute pointwise coverage rates of the 95\% credible intervals for $F_{\tmu}(y)$ on a grid of $y$ values, where the credible intervals are constructed using symmetric posterior quantiles. Additionally, we calculate pointwise bias, root mean squared error (RMSE), and credible interval lengths for $F_{\tmu}(y)$, averaging these quantities over data replicates for each  $y$. To summarize performance across the entire range of $y$, we compute overall metrics as weighted averages of these pointwise measures $m(y)$, where weights are given by the true baseline density $f_{\mu}$:
$$
M = \int_{\YY} m(y)\, f_{\mu}(y) \, dy = \mathbb{E}_{y \sim f_{\mu}}[m(y)].
$$
We refer to these summary measures as the \emph{weighted mean coverage}, \emph{weighted mean bias}, \emph{weighted mean RMSE}, and \emph{weighted mean interval length}. For exceedance probabilities, $p(y > y_0 \mid x)$, we obtain posterior mean estimates at selected quantile levels $y_0$, specifically at the $\alpha\%$-quantiles of the true conditional distribution $F_{y\mid x}$, with $\alpha \in \{10, 25, 50, 75, 90\}$. These estimates are computed for fixed covariate values $x=(1,0)$, $x=(1,0.25)$, and $x=(1,0.5)$. At each quantile $y_0$, we compute bias, RMSE, coverage rates of 95\% credible intervals, and credible interval lengths,  averaging these pointwise metrics across data replicates. Unlike the overall metrics calculated for $F_{\tmu}(y)$, these performance measures for exceedance probabilities are reported directly at the specified quantile points without any additional weighting.

Finally, for (Q3), we evaluate the frequentist bias of the regression coefficient estimates by computing their posterior means averaged across data replicates. To assess statistical efficiency in estimating the parameters $\beta_j$, we report both root mean squared error (RMSE) and lengths of the corresponding credible intervals. Additionally, we compare the performance of the proposed DPGLM to a parametric Beta regression model implemented via the \texttt{brms} (Bayesian Regression Models using Stan) package \citep{brms_package}. Specifically, in the Beta regression, both the mean response (with a logit link) and the precision parameter (with a log link) are modeled as functions of the covariates $x$. The comparison facilitates an evaluation of the more flexible semiparametric GLM structure in the DPGLM as compared to this parametric alternative. The data replicates used for this comparison are generated as described previously and not according to the Beta regression model.

\begin{figure}[!ht]
    \centering
    \includegraphics[width = 0.8\textwidth, height = 13.5cm]{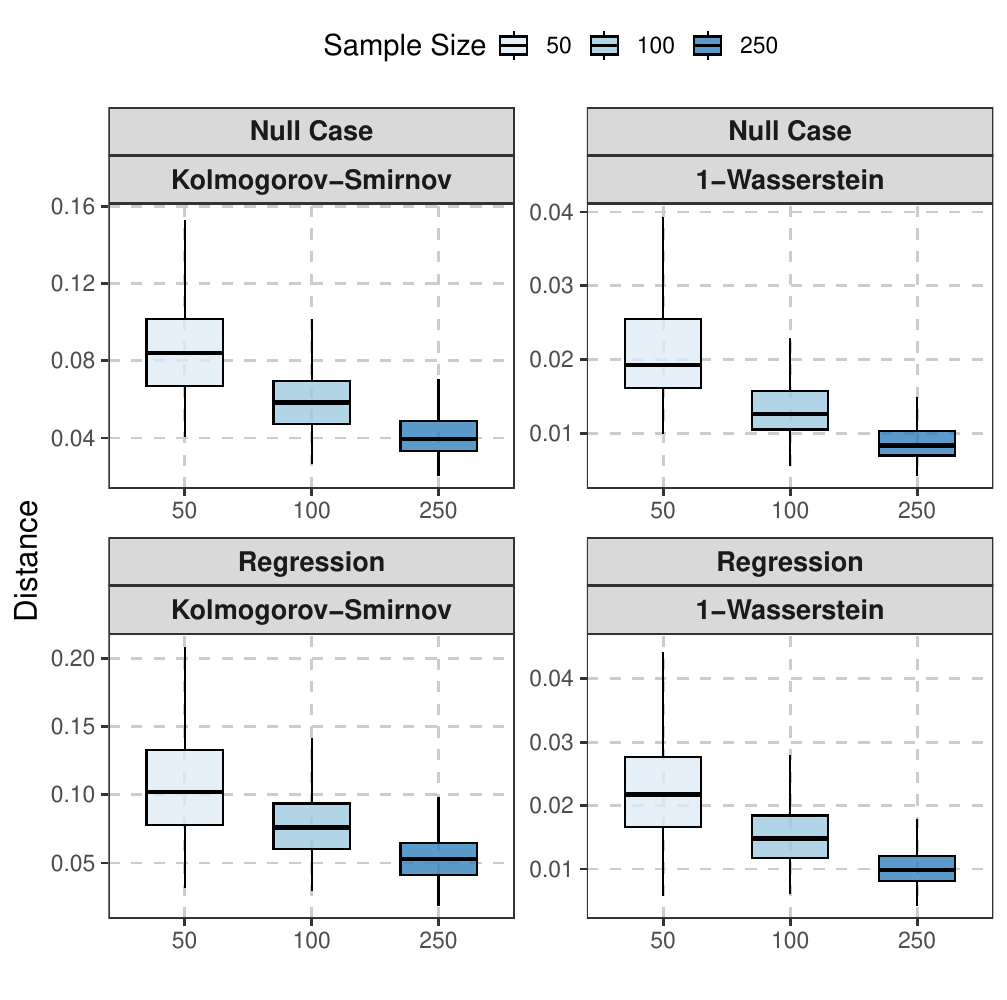}
    \caption{Kolmogorov--Smirnov (KS) statistics and 1-Wasserstein ($W_1$) distances for estimating the baseline cumulative distribution function, $F_{\tmu}(y)$, across varying sample sizes in the \emph{Null Case} (top panel) and \emph{Regression} simulation scenario (bottom panel). Results are based on \XX simulated datasets.}
    \label{fig:sim1}
\end{figure}

\begin{figure}[!ht]
    \centering
    \includegraphics[width = 0.75\textwidth, height = 9.7cm]{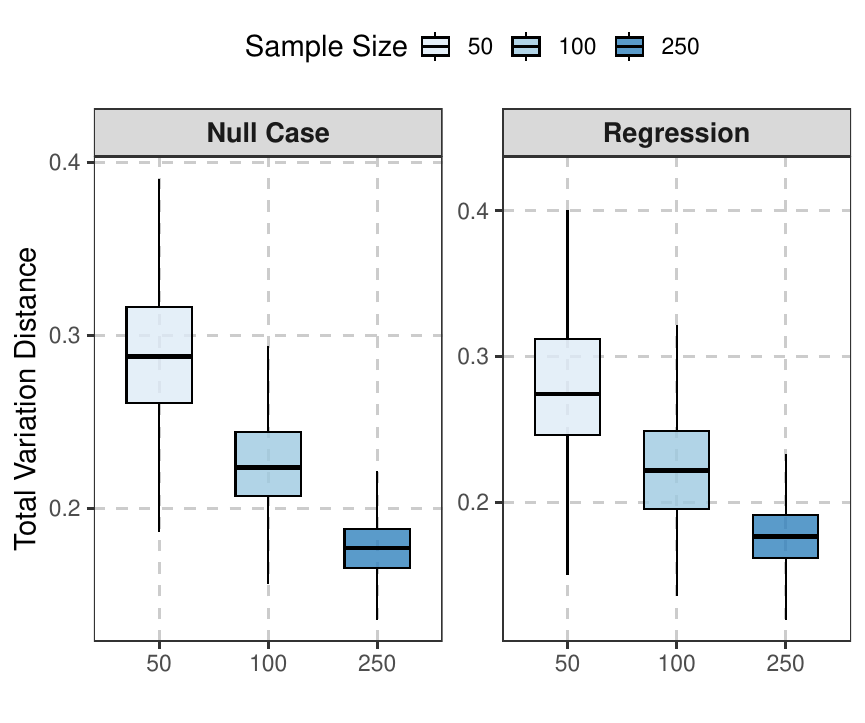}
    \caption{Total Variation (TV) distances for estimating the baseline density, $f_{\tmu}(y)$, across varying sample sizes in the \emph{Null Case} (left) and \emph{Regression} simulation scenario (right). Results are based on \XX simulated datasets.}
    \label{fig:sim2}
\end{figure}

\paragraph*{Results.} Figure~\ref{fig:sim1} presents boxplots (over \XX simulation replicates) of the Kolmogorov--Smirnov (KS) statistics and 1-Wasserstein ($W_1$) distances for estimating the baseline cumulative distribution function, $F_{\tmu}(y)$, across both simulation scenarios (\emph{Null Case} and \emph{Regression}) and varying sample sizes. Both KS and $W_1$ metrics decrease systematically with increasing sample size, demonstrating consistency of the DPGLM estimator for $F_{\tmu}(y)$. 
Figure~\ref{fig:sim2} displays the Total Variation (TV) distances for estimating the baseline density, $f_{\tmu}(y)$, across the same simulation scenarios. The TV distance consistently decreases with increasing sample size, indicating improved estimation accuracy for the baseline density $f_\tmu$.

\begin{figure}[!ht]
    \centering
    \includegraphics[width=0.9\textwidth, height = 11cm]{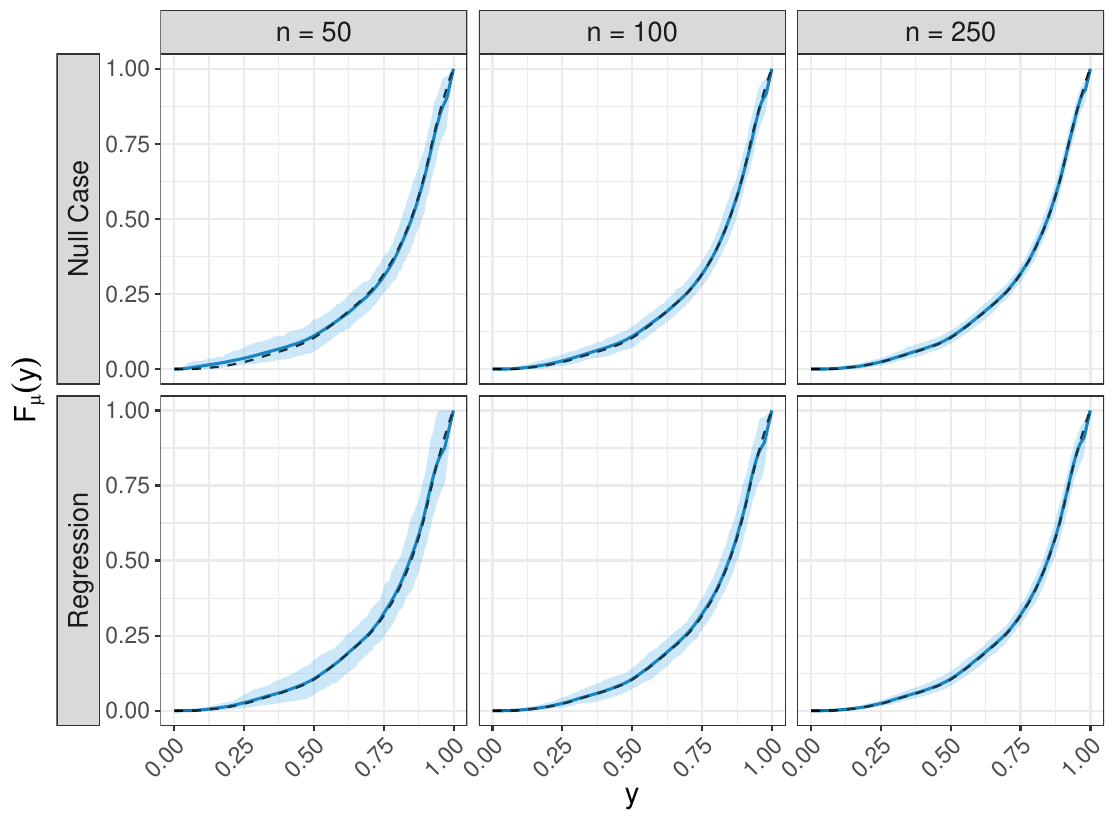}
    \caption{True (black dashed lines) and posterior mean estimates (blue solid lines) of $F_\tmu$ averaged over 200 simulated datasets. Results are shown for three sample sizes ($n=50$ on the left, $n=100$ in the middle, and $n=250$ on the right) and for two simulation scenarios: \emph{Null Case} (upper panel) and \emph{Regression} (lower panel). The shaded areas indicate the pointwise 2.5\% and 97.5\% quantile bands of the posterior estimates.}
    \label{fig:Fmu_estimates_cis}
\end{figure}

\begin{figure}[!ht]
    \centering
    \includegraphics[width = 0.82\textwidth, height = 13cm]{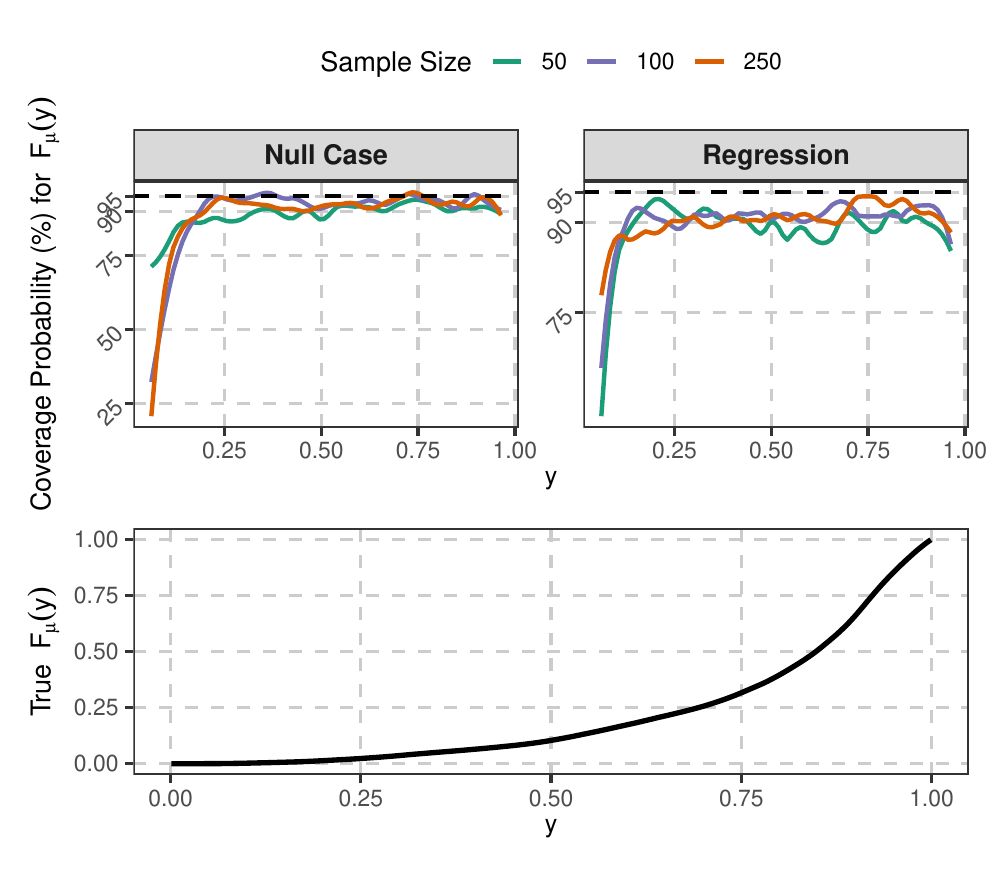}
    \caption{Pointwise coverage probabilities for the baseline CDF $F_{\tmu}(y)$, on a grid of $y$ values, for varying sample sizes, across both simulation scenarios: \emph{Null Case} (upper left) and \emph{Regression} (upper right). The black dashed line represents the nominal 95\% coverage level. The bottom panel displays the true $F_{\tmu}$. Results are based on \XX simulated data replicates.}
    \label{fig:sim4}
\end{figure}

Figure~\ref{fig:Fmu_estimates_cis} presents the true and posterior mean estimates of $F_{\tmu}$, averaged over data replicates, along with pointwise 95\% quantile bands. These bands become narrower and posterior means converge closer to the true distribution as the sample size increases. Pointwise coverage probabilities for $F_{\tmu}(y)$, presented in Figure~\ref{fig:sim4}, remain generally near the nominal 95\% level, with deviations observed at boundary regions (particularly close to $y=0$), are due to sparse data and simply reflect the lack of information (for example, there is no data in Speech Intelligibility below $y=0.12$). Table~\ref{tab:performance_metrics} reports the overall performance metrics for estimating $F_{\tmu}$, computed as weighted averages across the range of $y$ over simulated datasets. Both bias and RMSE decrease and credible intervals narrow with increasing sample size from $n=50$ to $n=250$, and coverage probabilities remain close to the nominal level. Similar trends under both Null and Regression scenarios indicate consistency and improved precision of the DPGLM estimates.

\begin{table}[!ht]
\centering
\caption{Overall performance metrics for $F_\tmu$, computed as weighted averages over the range of $y$ (using the true baseline density $f_\tmu$ as weights) across 200 simulated datasets. For each simulation scenario, we report the bias, RMSE, coverage probability (\%), and credible interval length at varying sample sizes.}
\vspace{0.3cm}
\begin{tabular}{llcccc}
\toprule
Scenario      & Sample Size ($n$) & {Bias}    & {RMSE}   & {Coverage (\%)} & {CI Length} \\
\midrule
Null Case     & 50          & -0.0065  & 0.0413   & 90.0           & 0.138   \\
              & 100         & -0.0025  & 0.0278   & 91.5           & 0.099  \\
              & 250         & -0.0003  & 0.0182   & 91.0           & 0.064  \\
\midrule
Regression    & 50          & 0.0019   & 0.0500   & 88.5           & 0.161   \\
              & 100         & 0.0012   & 0.0354   & 90.5           & 0.121   \\
              & 250         & 0.0004   & 0.0238   & 90.0           & 0.079  \\
\bottomrule
\end{tabular}
\label{tab:performance_metrics}
\end{table}

\begin{figure}[!ht]
    \centering
    \includegraphics[width = 0.85\textwidth, height = 13cm]{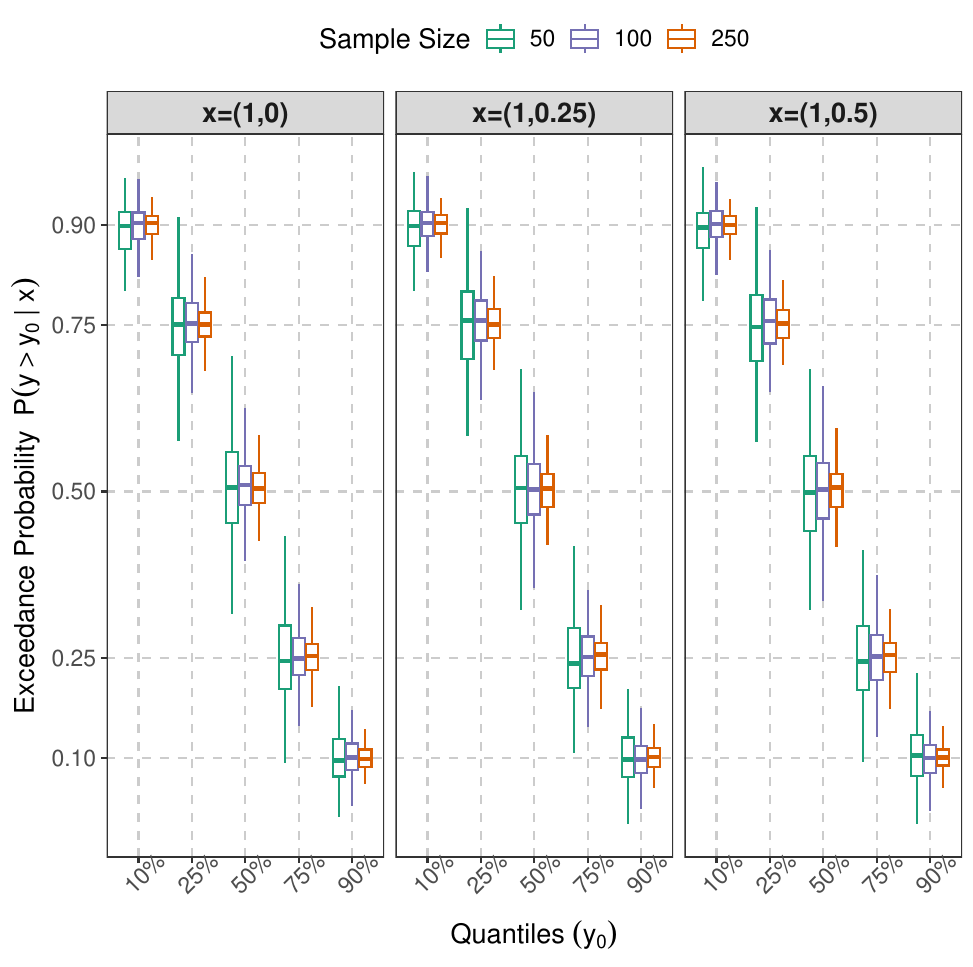}
    \caption{Posterior \emph{mean} estimates of exceedance probability, $p(y > y_0 \mid x)$, for $x = (1, 0)$ [\emph{left}], $x=(1, 0.25)$ [\emph{middle}], and $x = (1, 0.5)$ [\emph{right}], evaluated at various quantiles $y_0$ ($\alpha\%$-quantile of the true $F_{y\mid x}$, with $\alpha \in \{10, 25, 50, 75, 90\}$), based on \emph{regression} simulation scenario and \XX simulated data replicates.}
    \label{fig:exc_reg}
\end{figure}

\begin{table}[!ht]
    \centering
    \caption{Bias, Root-mean-square error (RMSE), coverage probability (\%), and CI length for the \emph{exceedance probability, $p(y > y_0 \mid x)$} at $x = (1, 0.5)$, evaluated at various quantiles $y_0$. Results are based on the second simulation scenario and \(\XX\) simulated data replicates.}
    \label{tab:exc_reg_3}
    \vspace{0.3cm}
    \begin{tabular}{l c c c c c c}
        \toprule
        \multirow{2}{*}{Metric} & \multirow{2}{*}{Sample Size ($n$)} & \multicolumn{5}{c}{Quantiles ($y_0$)}\\ 
        \cline{3-7} 
        &  & 10\% & 25\% & 50\% & 75\% & 90\% \\
        \midrule
        Bias 
            & 50  & -0.01 & 0.00 & 0.00 & 0.00 & 0.00 \\
            & 100 & 0.00  & 0.00 & 0.00 & 0.00 & 0.00 \\
            & 250 & 0.00  & 0.00 & 0.00 & 0.00 & 0.00 \\
        \midrule
        RMSE 
            & 50  & 0.04  & 0.07 & 0.09 & 0.07 & 0.05 \\
            & 100 & 0.03  & 0.05 & 0.06 & 0.05 & 0.03 \\
            & 250 & 0.02  & 0.03 & 0.03 & 0.03 & 0.02 \\
        \midrule
        Coverage (\%)
            & 50  & 94.5  & 92.5 & 92   & 92   & 87 \\
            & 100 & 91    & 92.5 & 90.5 & 91   & 91.5 \\
            & 250 & 95    & 95   & 94   & 93.5 & 93 \\
        \midrule
       CI Length 
            & 50  & 0.15  & 0.24 & 0.28 & 0.24 & 0.16 \\
            & 100 & 0.11  & 0.17 & 0.20 & 0.17 & 0.11 \\
            & 250 & 0.07  & 0.11 & 0.13 & 0.11 & 0.07 \\
        \bottomrule
    \end{tabular}
\end{table}

Table~\ref{tab:exc_reg_3} (with additional Tables~\ref{tab:exc_reg_1} and \ref{tab:exc_reg_2} provided in Appendix \ref{sec:additional_simulation}) and Figure~\ref{fig:exc_reg} summarize the estimation performance for exceedance probabilities, $p(y > y_0 \mid x)$, evaluated at covariate values $x=(1,0)$, $x=(1,0.25)$, $x=(1,0.5)$ and quantile levels (10\%, 25\%, 50\%, 75\%, and 90\%) of the true conditional distribution $F_{y\mid x}$. As the sample size increases from $n=50$ to $n=250$, posterior mean estimates become increasingly accurate, demonstrating reduced RMSE, narrower credible intervals, and negligible bias. Coverage probabilities remain close to the nominal 95\% level, although coverage at the highest quantile (90\%) at the smallest sample size $n=50$ is somewhat lower (e.g., 87\% for $x = (1, 0.5)$), reflecting sparse data near extreme values. Nevertheless, coverage notably improves with larger sample sizes, reaching approximately 93\% at $n=250$. These results illustrate the reliability and precision of the DPGLM in estimating exceedance probabilities, which are particularly relevant in clinical diagnostics and decision-making contexts.

\begin{table}[!ht]
\centering
\caption{Comparison of $\beta$ estimates across various sample sizes under simulation scenario I (\emph{null case}) for the proposed DPGLM approach and the Beta regression model using the \texttt{brms} package. Results are based on $B = 200$ simulation replicates.}
\vspace{0.3cm}
\begin{tabular}{l c c c c c c c c c}
\toprule
Pars & $n$ & \multicolumn{4}{c}{\textbf{DPGLM}} & \multicolumn{4}{c}{\textbf{Beta Regression}} \\ 
\cmidrule(lr){3-6} \cmidrule(lr){7-10}
     &   & Bias & RMSE & Coverage & CI Length & Bias & RMSE & Coverage & CI Length \\ 
\midrule
$\beta_0$ & 50  & 0.016  & 0.221  & 95.5   & 0.695   & -0.030   & 0.170   & 90.0   & 0.589   \\[0.5ex]
 & 100 & 0.008  & 0.109  & 96.5   & 0.445   & -0.034   & 0.115   & 94.0   & 0.410   \\[0.5ex]
 & 250 & 0.006  & 0.072  & 95.0   & 0.276   & -0.033   & 0.080   & 87.5   & 0.258   \\[1ex]
$\beta_1$ & 50  & 0.019  & 0.239  & 94.0   & 0.673   & 0.010    & 0.177   & 92.5   & 0.596   \\[0.5ex]
& 100 & 0.005  & 0.115  & 95.5   & 0.443   & 0.005    & 0.113   & 95.5   & 0.413   \\[0.5ex]
& 250 & 0.002  & 0.072  & 96.0   & 0.278   & 0.002    & 0.070   & 92.5   & 0.260   \\
\bottomrule
\end{tabular}
\begin{minipage}{\linewidth}
    \footnotesize
    \textbf{Notes:} True values are $\beta_0 = 1$ and $\beta_1 = 0$. Bias = Estimate - True value. Estimates, RMSE (Root Mean Square Error), Coverage Probability (\%), and CI Lengths are averaged over $B = 200$ data replicates. Abbreviations: n – Sample size; Pars – Parameters.
\end{minipage}
\label{tab:beta_null}
\end{table}

\begin{table}[!ht]
\centering
\caption{Comparison of $\beta$ estimates across various sample sizes under simulation scenario II (\emph{regression}) for the proposed DPGLM approach and the Beta regression model using the \texttt{brms} package. Results are based on \XX simulation replicates. True values are $\beta_0 = 0.2$ and $\beta_1 = 0.7$. For other details, see Notes in Table \ref{tab:beta_null}.}
\vspace{0.3cm}
\begin{tabular}{l c c c c c c c c c}
\toprule
Pars & n & \multicolumn{4}{c}{\textbf{DPGLM}} & \multicolumn{4}{c}{\textbf{Beta Regression}} \\ 
\cmidrule(lr){3-6} \cmidrule(lr){7-10}
     &   & Bias & RMSE & Coverage & CI Length & Bias & RMSE & Coverage & CI Length \\ 
\midrule
\(\beta_0\) & 50  & 0.007  & 0.171   & 92.0   & 0.592   & 0.007   & 0.145   & 96.0   & 0.562   \\[0.5ex]
 & 100 & 0.013  & 0.115   & 91.5   & 0.411   & 0.020   & 0.112   & 91.5   & 0.394   \\[0.5ex]
& 250 & 0.009  & 0.068   & 94.5   & 0.261   & 0.020   & 0.067   & 93.0   & 0.250   \\[1ex]
\(\beta_1\) & 50  & -0.011 & 0.155   & 94.0   & 0.579   & -0.064  & 0.160   & 91.5   & 0.571   \\[0.5ex]
 & 100 & -0.004 & 0.101   & 95.0   & 0.398   & -0.080  & 0.124   & 88.5   & 0.397   \\[0.5ex]
 & 250 & 0.002  & 0.061   & 96.5   & 0.256   & -0.081  & 0.101   & 77.0   & 0.249   \\
\bottomrule
\end{tabular}
\label{tab:beta_reg}
\end{table}

Tables~\ref{tab:beta_null} and~\ref{tab:beta_reg} compare regression parameter estimates obtained from the proposed DPGLM and a parametric Beta regression model (using the \texttt{brms} package) under both simulation scenarios. In the \emph{Null Case} scenario, the DPGLM produces nearly unbiased estimates for both $\beta_0$ and $\beta_1$, with bias decreasing from approximately $0.016$ at $n=50$ to $0.006$ at $n=250$ for $\beta_0$, and similarly for $\beta_1$. RMSE and credible interval lengths also substantially decrease with increasing sample size, indicating improved precision. Coverage probabilities for the DPGLM consistently approach the nominal 95\% level. In contrast, the Beta regression model, exhibits higher bias (about $-0.03$) and lower coverage (87.5\% at $n=250$) at larger samples, particularly for $\beta_0$. In the \emph{Regression} scenario, the DPGLM again demonstrates better performance. Bias for both parameters remains minimal and further reduces as sample size increases, while RMSE and credible interval lengths decline substantially. Coverage probabilities remain stable near the nominal level, around 92–96\%. The Beta regression model, however, demonstrates substantial bias in estimating $\beta_1$ (approximately $-0.08$), along with lower coverage rates (dropping to 77\% at $n=250$), suggesting potential estimation inconsistencies. These results highlight the flexibility, reliability, and better precision of the proposed DPGLM methodology. However, it is worth noting that these performance differences may partly reflect the fact that data were generated from an SPGLM (\citealp{rathouz2009generalized}) rather than a parametric Beta model; hence, scenarios where the data-generating mechanism aligns more closely with the Beta regression could yield more comparable performance.

\section{Application to Speech Intelligibility Data}
\label{sec:ex}
\begin{figure}[h]
    \centering
    \includegraphics[width = 0.95\textwidth, height = 10cm]{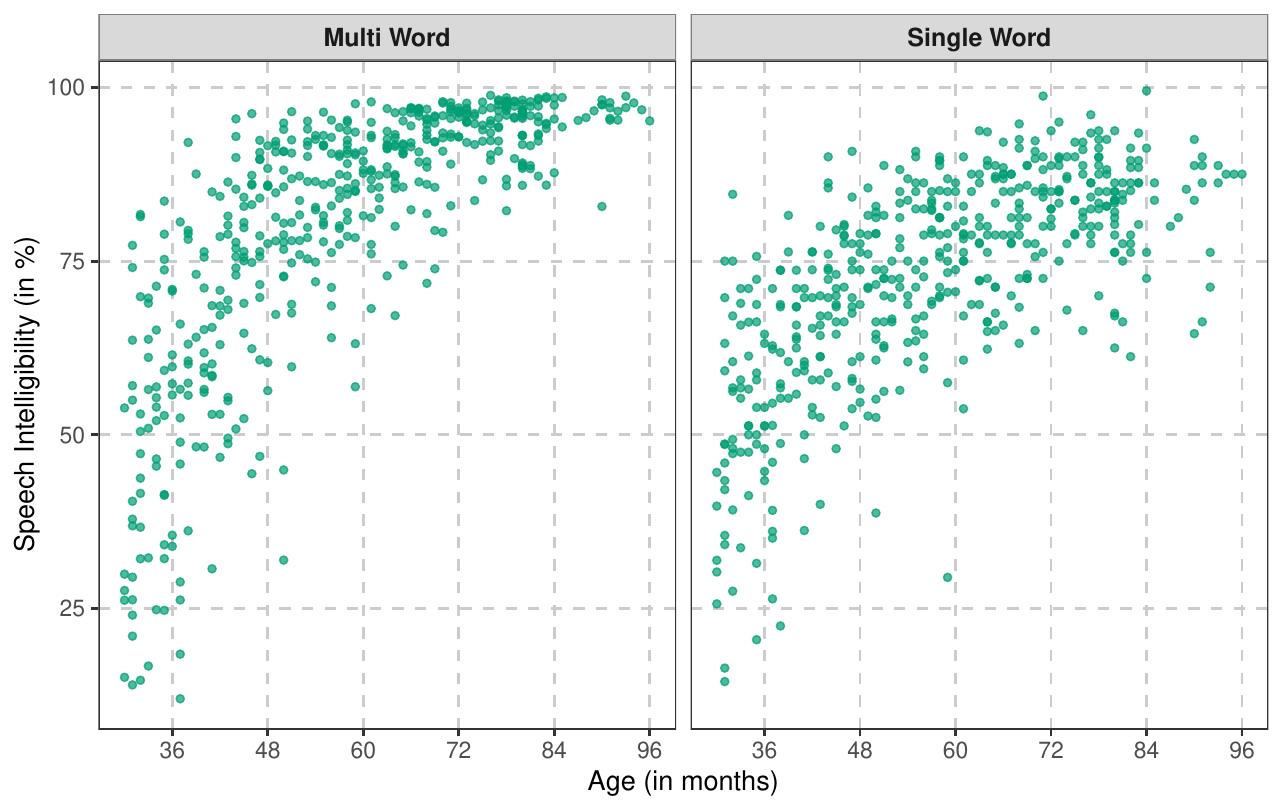}
    \caption{Observed data. Mean speech intelligibility (in percentage) for single-word (SW, right panel) and multi-word (MW, left panel) utterances, with age (in months) as a predictor.}
    \label{fig:1}
\end{figure}

We implement inference under the DPGLM  for a data set from a speech intelligibility study for typically developing (TD) children from
$30$ to $96$ months of age (see Figure \ref{fig:1}). The study included $n = 505$ typically developing children,
wherein \textit{mean speech intelligibility} was measured as the
proportion of words correctly transcribed by two naive evaluating adult
listeners. The \textit{mean speech intelligibility} (MSI) was recorded
separately for \textit{single-word} and \textit{multi-word}
utterances, and we refer them as SW-MSI and MW-MSI, respectively. For
further details on the dataset, we refer to
\citet{mahr:2020} and \citet{hustad:2021}.

We carry out two separate analyses for the SW- and MW-MSI, respectively. In both cases, MSI is the response $y_i$ for child $i$, $\ i=1,\ldots,n$. Covariates $x_i$ are defined to allow for a non-linear regression of $y_i$ on age. We use the basis functions of a 3-df natural cubic spline to model MSI as a function of age. Note that these 3 degrees of freedom pertain only to a mean zero spline basis; the intercept constitutes an additional (fourth) degree of freedom. 
This allows the mean to vary flexibly with age. We use a logit link in the GLM regression. Next, considering a uniform$(z-c, z+c)$ kernel $K$ with $c$ proportional to Silverman's rule of thumb \citep{silverman1986} and the prior distributions as in (\ref{DPGLM}) with $\alpha = 1$ and $G_0$ = uniform$(0,1)$, we fit the proposed DPGLM model to the speech intelligibility study using the MCMC algorithm detailed in Appendix \ref{sec:mcmc}, generating a total of $3,000$ MCMC samples. We discard the first $1,000$ iterations as initial burn-in and use the remaining $R = 400$ Monte Carlo samples, after thinning by a factor of $5$, for the following results. 

\begin{figure}[!ht]
    \centering
    \includegraphics[width = 0.95\textwidth, height = 10.5cm]{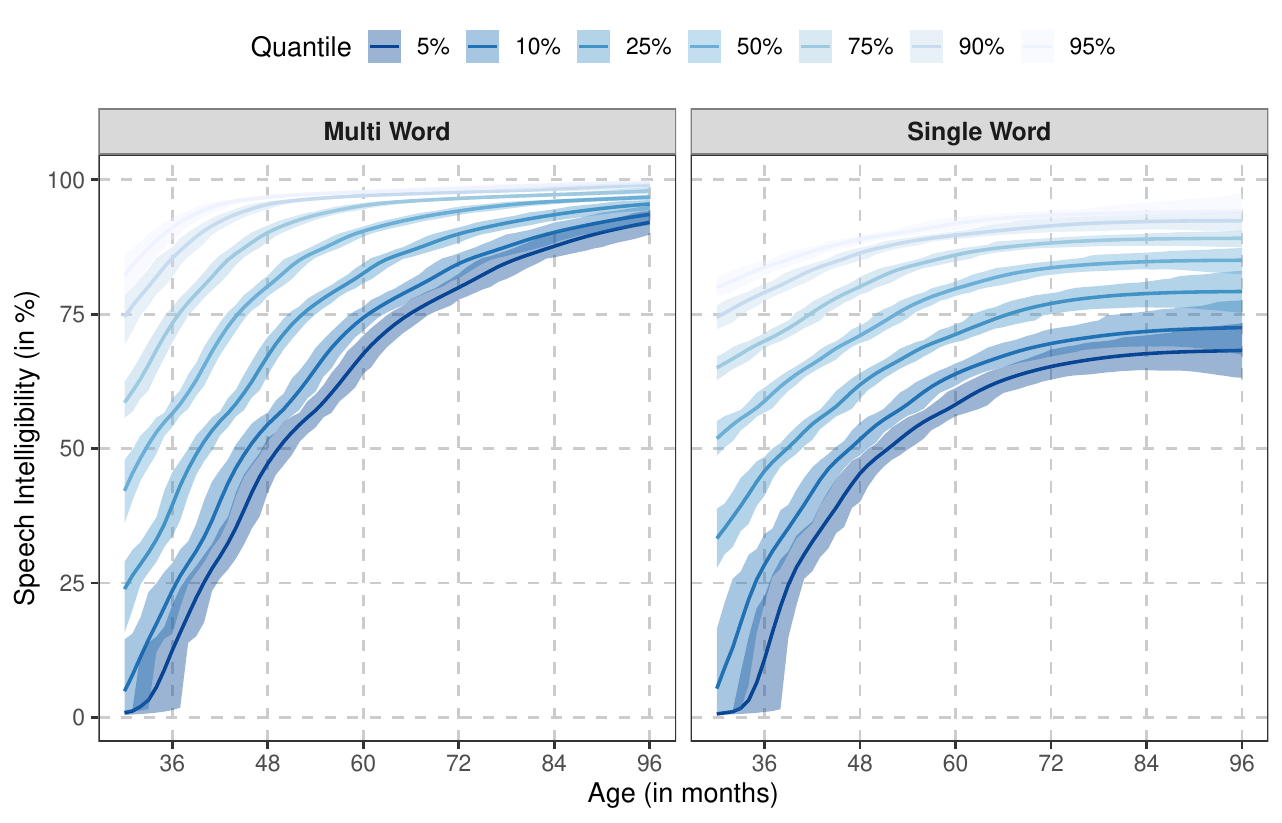}
    \caption{Quantile growth curves (solid lines) based on DPGLM model for multi-word (left panel) and single-word (right panel) intelligibility, with $95\%$ point-wise uncertainty intervals (shaded ribbon).}
    \label{fig:2}
\end{figure}

\paragraph*{Results.} Figure \ref{fig:2} illustrates the extracted quantile regression curves, $q_{\alpha}(\bx)$, based on the proposed model for single-word and multi-word intelligibility, accompanied by $95\%$ point-wise uncertainty intervals. The curves represent various quantiles ($\alpha = 5\%, 10\%, 25\%, 50\%, \\ 75\%, 90\%,$ and $95\%$) of speech intelligibility as a function of age in months, indicating that intelligibility improves with age. The wider uncertainty intervals at younger ages reflect greater variability, underscoring the model's effectiveness in capturing nuances of speech development and providing valuable insights for pediatric speech-language pathology.

\begin{figure}[!ht]
    \centering
    \includegraphics[width = 0.95\textwidth, height = 10cm]{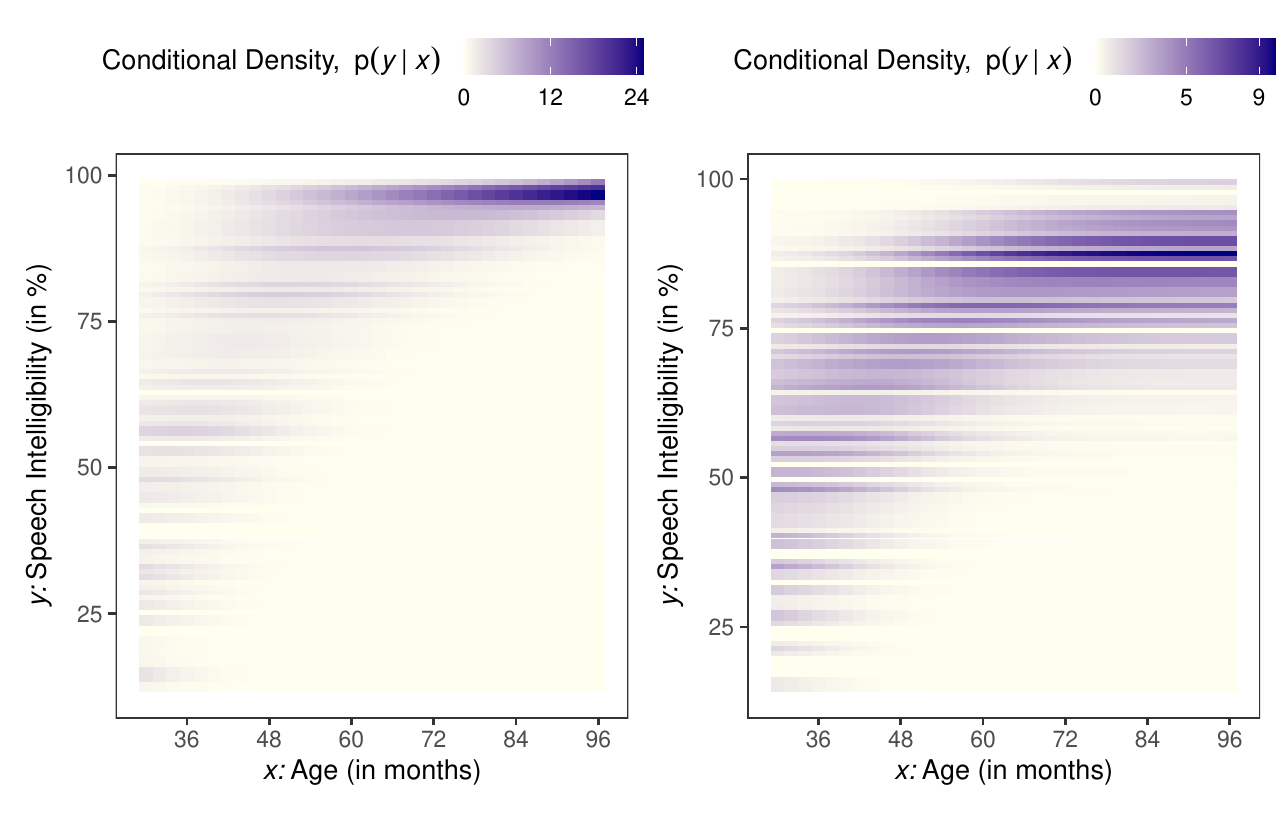}
    \caption{Heatmap for fitted densities, $\widehat p(y \mid x)$ (read vertically), with speech intelligibility (in percentage) as response $y$ and age (in months) as covariate $x$, for multi-word (left panel) and single-word (right panel) utterances. Here the gradient from \textit{ivory} to \textit{navy} represents an increase in $p(y \mid x)$.}
    \label{fig:3}
\end{figure}

Figure \ref{fig:3} presents the fitted densities, $\widehat p(y \mid x)$ on a grid of age values running along the horizontal, illustrating the relationship between speech intelligibility (in percentage) as the response variable $y$ and age as the covariate $x$. The heatmap shows how these densities vary across ages, with a gradient from white to blue indicating increased $\widehat p(y \mid x)$. This visualization complements the quantile regression analysis from Figure \ref{fig:2} by reinforcing the trend that older children achieve higher intelligibility scores. 

\begin{figure}[!ht]
    \centering
    \includegraphics[width = 0.95\textwidth, height = 10.5cm]{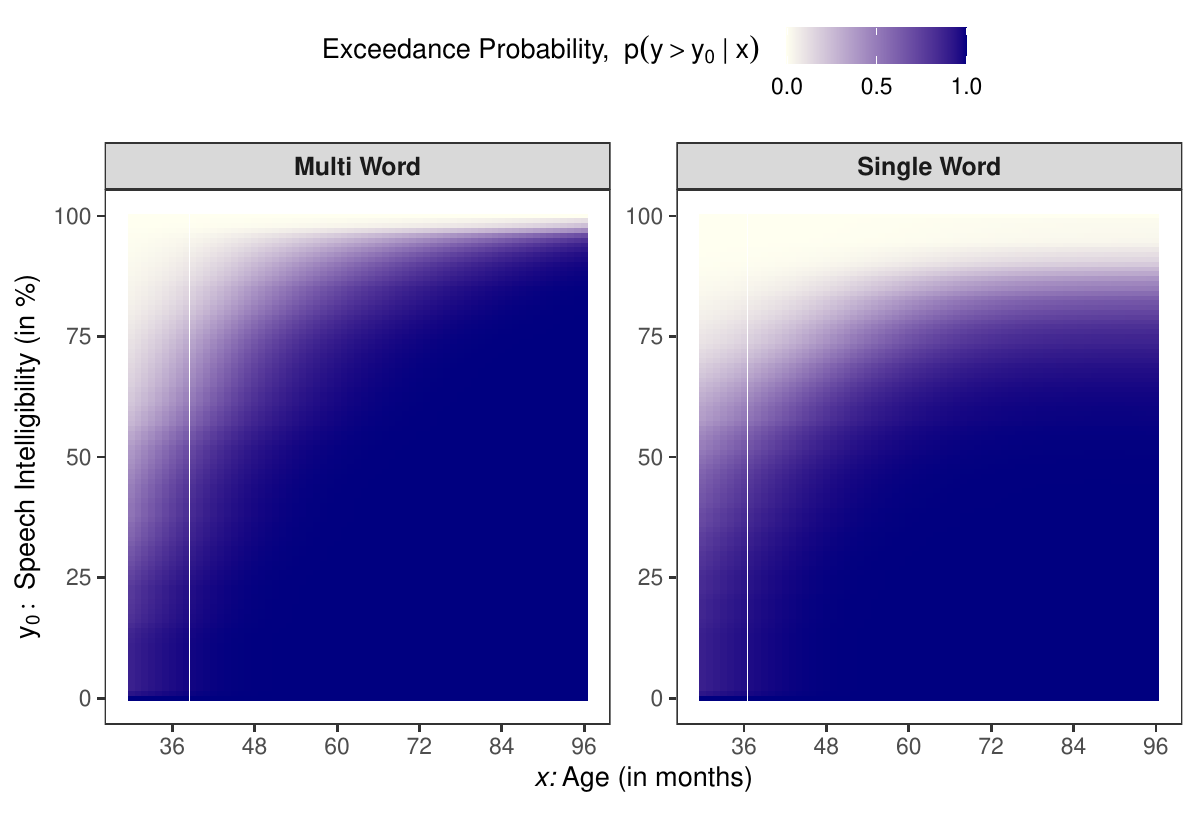}
    \caption{Estimate for exceedance probabilities, $p(y > y_0 \mid x)$ (read vertically), with speech intelligibility (in percentage) as response threshold $y_0$ and age (in months) as covariate $x$. Here the gradient from \textit{ivory} to \textit{navy} represents an increase in $p(y > y_0 \mid x)$.}
    \label{fig:4}
\end{figure}

\begin{figure}[!ht]
    \centering
    \includegraphics[width = 0.85\textwidth, height = 9.5cm]{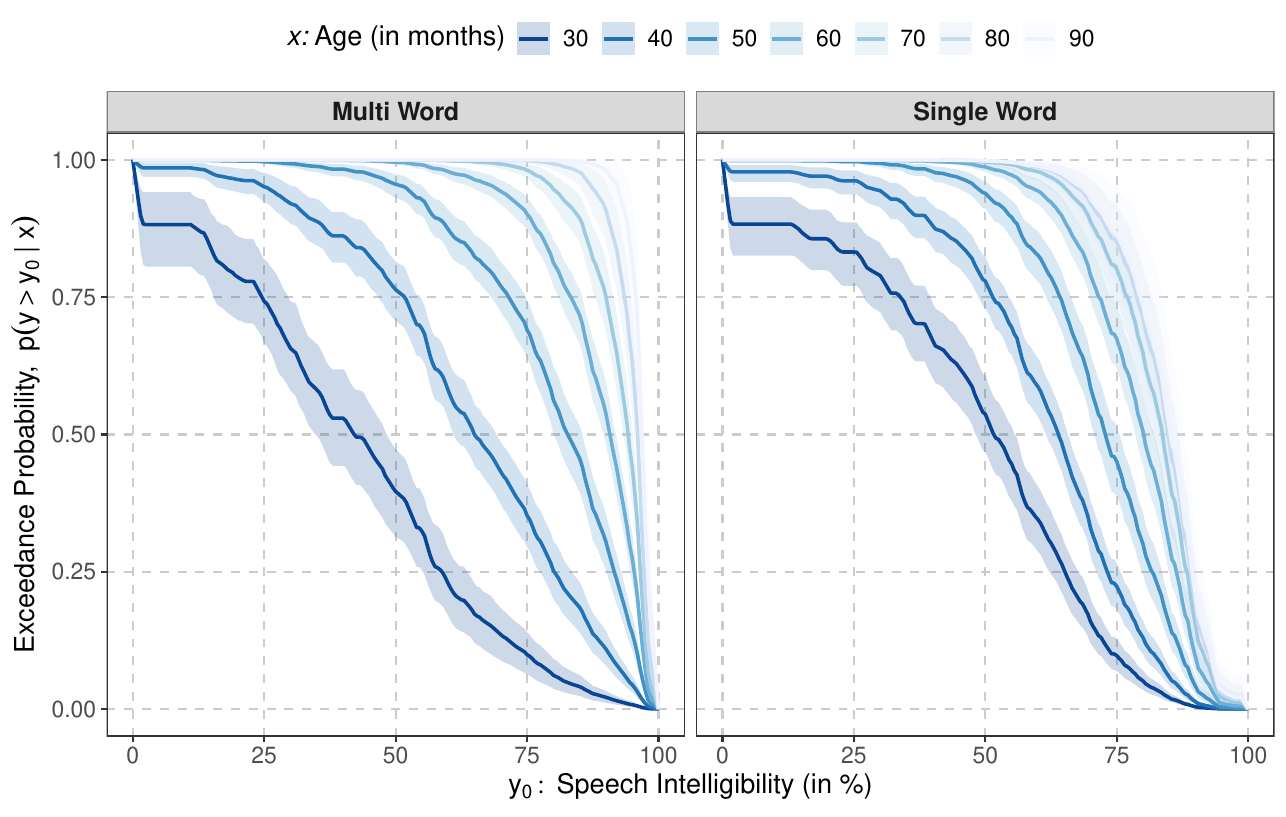}
    \caption{Estimate for exceedance probabilities, $p(y > y_0 \mid x)$, with $95\%$ point-wise uncertainty intervals for varying ages. Here speech intelligibility (in percentage) is considered as response threshold $y_0$ and age (in months) as covariate $x$.}
    \label{fig:5}
\end{figure}

Figure \ref{fig:4} displays estimates for exceedance probabilities, $p(y > y_0 \mid x)$, across varying ages $x$ (horizontal axis) and thresholds $y_0$ (vertical axis), indicating the likelihood that speech intelligibility $y$ exceeds a threshold $y_0$. Figure \ref{fig:5} enhances this analysis by incorporating 95\% point-wise uncertainty intervals, providing a clearer understanding of variability in exceedance probabilities at different ages $x$ (color shades). Together, these visualizations underscore developmental trends in speech intelligibility, highlighting that older children are more likely to achieve higher levels of intelligibility. More importantly, for this methodological development, these results show how, once fitted, our Bayesian implementation of the SPGLM can produce inferences on a variety of useful derived model parameters.  

\section{Discussion}
\label{sec:disc}
We have introduced an extension of the GLM family for continuous response data to a semiparametric model with
a BNP prior on the baseline distribution.  Using a NRM with homogeneous Levy intensity as prior model we characterized the posterior distribution using an inhomogeneous NRM. While NRM priors, in particular, the special
case of DP priors, are widely used in BNP inference, only few applications naturally give rise to inhomogeneous NRM's.
It is interesting to note that this naturally happens with the exponential tilting in the GLM model.

One of the limitations of the model is the restricted structure implied by the GLM framework which assumes that the sampling model indexed by different covariates changes only by exponential tilting of the same underlying baseline distribution. While the parsimony of this structure is often desirable, it is also introducing a limitation, by making it more difficult to model certain scenarios. For example, if the sampling model were to include multimodality for extreme values of the covariate, as it might happen for some clinical outcomes, this would be more naturally modeled with more flexible dependent DP models, and difficult to capture with the proposed DPGLM.

Several extensions and generalizations of the proposed model could be considered,
including extension to multivariate outcomes and for repeated measurements. The latter could include subject specific random effects. For inference with latent variables such as random effects, Bayesian inference is typically more natural and allows easier implementation than, say, maximum likelihood estimation, which can require onerous numerical integration and poses greater challenges in extracting derived parameters, such as marginal (over random effects) trends.
Finally, we believe that the Bayesian DPGLM could be an attractive option in data analysis for clinical studies, including planning and sample size arguments for future studies. One particular advantage is the straightforward inference for any desired summary or function of the unknown quantities. One can report inference or plan study designs with focus on any clinically relevant summary, such as exceedance probabilities etc.

\subsubsection*{Acknowledgement}
The authors thank Igor Pr\"unster (personal communication) for advice on the proof of Proposition \ref{result1}, and for pointing us to the cited result.  The authors also thank Katherine Hustad and her very able team for sharing the speech intelligibility data used in Section~\ref{sec:ex}. 

\subsubsection*{Funding}
This research was supported by the National Institutes of Health (NIH) under grants 2R01 HL094786 (J. Schildcrout, PI) and R01DC015653 (K. Hustad, PI).

\bibliography{references}

\newpage
\appendix
\section*{Appendix}
\section{Proofs}
\label{sec:proofs}
We include proofs for Propositions \ref{result1} through \ref{result3}. The proof for Proposition \ref{result1} is summarized from \citet{nieto2004normalized}. 

\paragraph*{Proof of Proposition \ref{result1}.}
Let $\tmu$ denote a gamma completely random measure (CRM) with base measure $G_0$ on support $\YY$ and concentration parameter $\alpha$. In particular, $\tmu$ can be represented as $\tmu(\cdot)=\sum_{\ell=1}^{\infty} s_\ell \delta_{z_\ell}(\cdot)$ with Levy intensity
$\nu(ds, dz)= \rho(ds) \, \alpha G_0(dz) = \frac{e^{-s}}{s} ds \, \alpha G_0(dz)$. It follows that $\tmu(\YY) \sim \text{Gamma}(\alpha, 1)$ and the normalized measure $\frac{\tmu}{\tmu(\YY)} := f_{\tmu} \sim \text{DP}(\alpha, G_0)$. Under DPGLM, the implied marginal $\Gx(z)$ for a given $x$ is $\Gx(z) \propto \exp(\theta_x z)\tmu(z)$. For a fixed $\theta_x$, the exponentially tilted CRM can be written as
$
\tmu^\star = \exp(\theta_x z) \tmu(z) \\ = \sum_{\ell=1}^{\infty} s^\star_\ell \delta_{z_\ell}
$
with $s^\star_\ell= \exp \left(\theta_x z_\ell\right) s_\ell$. 

Note that for any measurable function $g$, $\E\left\{e^{-\int_\YY g(z) \tmu^\star(d z)}\right\} =  \E \left\{e^{-\int_\YY g^\star(z) \tmu(d z)}\right\}$, where $g^\star(z) = \exp(\theta_x z) g(z)$. Since $g$ is arbitrary and measurable, so is $g^\star$. Using the Levy-Khintchine representation for $\tmu$, we have 
$$
\E \left\{e^{-\int_\YY g^\star(z) \tmu(d z)}\right\} = \exp \left[-\int_{\RR^{+} \times \YY}\left\{1-e^{-s g^\star(z)}\right\} \nu(ds, dz)\right].
$$
With the change of variable $s\to s^\star$, where $s^\star=s\,\exp(\theta_x z)$, this expression becomes
$$
\exp \left[-\int_{\RR^{+} \times \YY}\left\{1-e^{-s g^\star(z)}\right\} \nu(ds, dz)\right] =  \exp \left[-\int_{\RR^{+} \times  \YY}\left\{1-e^{-s^\star g(z)}\right\} \frac{e^{-s^\star/\exp(\theta_x z)}}{s^\star} ds^\star \, \alpha G_0(dz) \right]
$$
Thus, the Levy-Khintchine representation for $\tmu^\star$ is
$$
\E\left\{e^{-\int_\YY g(z) \tmu^\star(d z)}\right\} = \exp \left[-\int_{\RR^{+} \times  \YY}\left\{1-e^{-s g(z)}\right\} \nu^\star(ds, dz) \right],
$$
with Levy intensity 
$$
\nu^\star(ds, dz) = \frac{e^{-s/\exp(\theta_x z)}}{s} ds \, \alpha G_0(dz) = \rho^\star(ds \mid z) \, \alpha G_0(dz),
$$ 
where $\rho^\star(ds\mid z)=\frac{e^{-s/\exp(\theta_x z)}}{s}\,ds$ depends on the atom location $z$, thereby characterizing a non-homogeneous CRM. This, in turn, implies that $\Gx$ is a non-homogeneous normalized random measure (NRM), which completes the proof.

\paragraph*{Proof of Proposition \ref{result3}.} 
Throughout the proof we treat $\th_x = \th_x(\tmu)$ as fixed. We start by defining $T_i \equiv T_i(\sy) = T(\th_{x_i}) = \exp\{b(\th_{x_i})\}$ with $b(\theta_x)$ given in (\ref{eq:2}). For simplicity, we consider the case of no ties in $\{z_i\}^n_{i=1}$, which is extended to the general case later. Let $C_i := \{z \in \sy: d(z, z_i) < \eps\}, \, i = 1, \dots, n,$ be $n$ disjoint subsets of $\sy$, where $d$ is a distance function, and define $C_{n+1} = \sy \setminus \cup^n_{i=1} C_i$. Denote $\widetilde T_i = T_i(C_i) =  \int_{C_i} \exp (\th_{x_i} z)\tmu(dz)$. We then obtain
\begin{align}
\label{sup:eq1}
\E \left(e^{-\int_\sy h(z) \tmu(dz)} \mid z_1 \in C_1, \dots, z_n \in C_n\right) & =  
\frac{\int e^{-\int_\sy h(z) \tmu(dz)} \prod_{i=1}^n \frac{T_i(C_i)}{T_i(\sy)} p(\tmu) d(\tmu)}{\int \prod_{i=1}^n \frac{T_i(C_i)}{T_i(\sy)} p(\tmu) d(\tmu)} \nonumber \\
& =   \frac{\E_{\tmu}\left( e^{-\int_\sy h(z) \tmu(dz)} \prod_{i = 1}^n \frac{\widetilde T_i}{T_i} \right)}{\E_{\tmu}\left(\prod_{i=1}^n \frac{\widetilde T_i}{T_i} \right)} = \frac{\E_{\tmu}(N)}{\E_{\tmu}(D)} \ .
\end{align}
Our goal is to derive the Laplace functional of the posterior for $\tmu$ (with fixed $\bth$) by letting $\eps \to 0$. First, note that setting $h(z) = 0$ in (\ref{sup:eq1}) yields the denominator. Next, using $\frac{1}{T_i} = \int_{\R^{+}} e^{-T_i u_i} d u_i =  \int_{\R^{+}} e^{- \int_{\mathcal{Y}} u_i \exp (\th_i z) \tmu(dz)} d u_i$, we obtain
$$
\prod_{i=1}^n \frac{1}{T_i} = \int_{{\R^{+}}^n} \prod_{j=1}^{n+1} e^{- \int_{C_j} \sum_{i=1}^n u_i \exp (\th_i z) \tmu(dz)} d \ub \ .
$$
Writing the numerator $N$ explicitly, we have
\begin{align*}
   & =  \prod_{j=1}^{n+1} e^{-\int_{C_j} h(z) \tmu(dz)}  \prod_{i=1}^n {\widetilde T_i} \int_{{\R^{+}}^n} \prod_{j=1}^{n+1} e^{- \int_{C_j} \sum_{i=1}^n u_i \exp (\th_i z) \tmu(dz)} d \ub \\
   & =  \prod_{i=1}^n {\widetilde T_i} \int_{{\R^{+}}^n} \prod_{j=1}^{n+1} e^{- \int_{C_j} \left(h(z) + \sum_{i=1}^n u_i \exp (\th_i z) \right) \tmu(dz)} d \ub  \\
  & =  \int_{{\R^{+}}^n}  e^{- \int_{C_{n+1}} \left(h(z) + \sum_{i=1}^n u_i \exp (\th_i z) \right) \tmu(dz)} \prod_{j=1}^{n} \left\{ - \frac{d}{du_j} e^{- \int_{C_j} \left(h(z) + \sum_{i=1}^n u_i \exp (\th_i z) \right) \tmu(dz)} \right\} d \ub 
\end{align*}
The last equality follows from $\frac{d}{du_j} e^{- \int_{C_j} \left(h(z) + \sum_{i=1}^n u_i \exp (\th_i z) \right) \tmu(dz)} = - \widetilde {T_j} e^{- \int_{C_j} \left(h(z) + \sum_{i=1}^n u_i \exp (\th_i z) \right) \tmu(dz)}$. Choosing $\epsilon$ sufficiently small so that $\tmu(C_j)$ are independent across $j$ and by Fubini's theorem, we get 
$$
\E_{\tmu}(N) = \int_{{\R^{+}}^n}  \E_{\tmu}\left[e^{- \int_{C_{n+1}} \left(h(z) + \sum_{i=1}^n u_i \exp (\th_i z) \right) \tmu(dz)} \right] \prod_{j=1}^{n} - \frac{d}{du_j} \E_{\tmu}\left[e^{- \int_{C_j} \left(h(z) + \sum_{i=1}^n u_i \exp (\th_i z) \right) \tmu(dz)} \right] d \ub  
$$
Introducing the notation 
$$
\eta(z, \ub) := h(z) + \sum_{i=1}^n u_i \exp (\th_i z)
$$ and writing $S = \R^{+} \times \sy = \cup_{j=1}^{n+1} S_j$ with $S_j = \R^{+} \times C_j$. Using the L\'evy–Khintchine representation and $S_{n+1} = S \setminus \cup_{j=1}^n S_j$, 
\begin{align*}
  \E_{\tmu}(N) & = \int_{{\R^{+}}^n}   \exp \left \{ - \int_{S_{n+1}} \left[1 - e^{- s \eta(z, \ub)} \right] \nu(ds, dz)\right\}  \prod_{j=1}^{n} - \frac{d}{du_j}  \exp \left \{- \int_{S_j} \left[1 - e^{- s \eta(z, \ub)} \right] \nu(ds, dz)\right\} d \ub  \\
 & =  \int_{{\R^{+}}^n}   \exp \left \{ - \int_{S} \left[1 - e^{- s \eta(z, \ub)} \right] \nu(ds, dz)\right\} \,  \prod_{j=1}^{n} V^{(1)}_{C_j}(\ub)  d \ub  \ ,
 \end{align*}
with $V^{(1)}_{C_j}(\ub) = \left\{- \frac{d}{du_j}  \exp \left( - \int_{S_j} \left[1 - e^{- s \eta(z, \ub)} \right] \nu(ds, dz)\right) \right\} \, \exp \left \{ \int_{S_j} \left[1 - e^{- s \eta(z, \ub)} \right] \nu(ds, dz)\right\}$. 

\noindent Using $\nu(ds, dz) = \rho(ds \mid z) \, \alpha G_0(dz)$ from (\ref{DPGLM}), we have
\begin{align*}
   V^{(1)}_{C_j}(\ub) & =  - \frac{d}{du_j}  \left( - \int_{S_j} \left[1 - e^{- s \eta(z, \ub)} \right] \nu(ds, dz)\right) \\
    & =  \int_{C_j} \exp (\th_j z)  \left\{\int_{\R^{+}} s e^{- s \eta(z, \ub)} \  \rho(ds \mid z) \right\} \alpha G_0(dz) \\
   & = \int_{C_j} \exp (\th_j z)  \phi_{1}(\ub, z) V^{(0)}_{C_j}(\ub)\, \alpha G_0(dz) =  \int_{C_j}  \exp (\th_j z)  \phi_{1}(\ub, z) \, \alpha G_0(dz),
\end{align*}
with $V^{(0)}_{C_j}(\ub) = 1$ and $\phi_{1}(\ub, z) = \int_{\R^{+}} s e^{-s\left(h(z) + \sum_{i=1}^n u_i \exp(\th_i z)\right)} \rho(ds \mid z)$. Denoting $\Delta^{(1)}_{\th_j} (\ub, z) := \exp (\th_j z)  \phi_{1}(\ub, z)$, and noting that the denominator $\E_{\tmu}\left[\prod_{i = 1}^n \frac{\widetilde T_i}{T_i}\right]$ in (\ref{sup:eq1}) is the same as the numerator with $h(z)=0$ (i.e., replacing $\eta(z, \ub)$ by $\eta^\star(z, \ub) =  \sum_{i=1}^n u_i \exp (\th_i z)$), we obtain
\begin{align}
\label{sup:eq2}
  & \E_{\tmu} \left(e^{-\int_\sy h(z) \tmu(dz)} \mid z_1 \in C_1, \dots, z_n \in C_n\right) \nonumber \\
  & =  \frac{\int_{{\R^{+}}^n}   \exp \left[ - \int_{S} \left\{1 - e^{- s\eta(z, \ub)} \right\} \nu(ds, dz) \right] \,  \prod_{j=1}^{n}   \int_{C_j}   \Delta^{(1)}_{\th_j} (\ub, z) \, \alpha G_0(dz) d \ub  }{\int_{{\R^{+}}^n}   \exp \left[ - \int_{S} \left\{1 - e^{- s \eta^\star(z, \ub)} \right\} \nu(ds, dz) \right] \,  \prod_{j=1}^{n}   \int_{C_j}  \Delta^{\star (1)}_{\th_j} (\ub, z) \, \alpha G_0(dz) d \ub }, 
\end{align}
with $\Delta^{\star (1)}_{\th_j} (\ub, z) = \exp (\th_j z)   \int_{\R^{+}} s e^{-s \eta^\star(z, \ub)} \rho(ds \mid z)$. Upon pushing $\epsilon \to 0$, we get 
$$
\int_{C_j}  \Delta^{\star (1)}_{\th_j} (\ub, z) \, \alpha G_0(dz) \to \alpha \exp (\th_j z_j)   \int_{\R^{+}} s e^{-s\eta^\star(z_j, \ub)} \rho(ds \mid z_j),
$$
and similarly, $\int_{C_j}  \Delta^{(1)}_{\th_j} (\ub, z) \, \alpha G_0(dz) \to \alpha \exp (\th_j z_j) \int_{\R^{+}} s e^{-s\eta(z_j, \ub)} \rho(ds \mid z_j)$.
A decomposition of $\left\{1 - e^{- s\eta(z, \ub)} \right\}$ shows that 
$$
\exp \left[ - \int_{S} \left\{1 - e^{- s\eta(z, \ub)} \right\} \nu(ds, dz) \right] = \exp \left[ - \int_{S} \left\{1 - e^{- h(z) s} \right\} \nuc(ds, dz) \right] \, e^{-\psi(\ub)},
$$
where $\nuc(ds, dz) = e^{- \eta^\star(z, \ub) s}\nu(ds, dz)$ and $e^{-\psi(\ub)} = e^{\left[ - \int_{S}  \left\{1 - e^{- \eta^\star(z, \ub) s} \right\} \nu(ds, dz) \right]}$. Letting $\zb = (z_1, \dots, z_n)$ and substituting everything in (\ref{sup:eq2}), we get
\begin{align}
\label{sup:eq3}
  & \E_{\tmu} \left(e^{-\int_\sy h(z) \tmu(dz)} \mid \zb \right) \nonumber \\
  & = \frac{\int_{{\R^{+}}^n} \E_{\muc}\left\{e^{-\int_\sy h(z) \muc (dz)}\right\} \left[\prod_{j=1}^n \E_{J_j}\left\{e^{-h(z_j) J_j} \right\}\right] D(\ub) e^{-\psi(\ub)} d \ub }{\int_{{\R^{+}}^n} D(\ub) e^{-\psi(\ub)} \ d \ub}, 
\end{align}
where $D(\ub) = \prod_{j=1}^n \int_{\R^{+}} s e^{-s \eta^\star(z_j, \ub)} \rho(ds \mid z_j)$. For fixed $\bth$ (by a slight abuse of notation, we include $\thb$ in the conditioning subset, to indicate the fixed $\thb$), writing $\tmu^\star \stackrel{d}{=} \tmu \mid \zb, \ub, \bth \stackrel{d}{=} \muc + \sum_{j=1}^n J_j \delta_{z_j}$, Eq. (\ref{sup:eq3}) can be expressed as
\begin{align}
\label{sup:eq4}
  \E_{\tmu} \left(e^{-\int_\sy h(z) \tmu(dz)} \mid \zb \right) = \int_{{\R^{+}}^n} \E_{\tmu}\left\{e^{-\int_\sy h(z) \tmu (dz)} \mid \zb, \ub \right\} \frac{D(\ub) e^{-\psi(\ub)}}{\int_{{\R^{+}}^n} D(\ub) e^{-\psi(\ub)} \ d \ub} \ d \ub, 
\end{align}
where $\muc \sim \CRM\left(\nuc\right)$ with L\'evy intensity  
$$
\nuc(ds, dz) = e^{- \eta^\star(z, \ub) s}\nu(ds, dz) = \frac1s\, {e^{-s\left\{1 + \sum_{i=1}^n u_i \exp(\th_i z) \right\}}} ds \, \alpha G_0(dz),
$$ and $P_{J_j}\left(s \mid z_j, \ub\right) \propto s e^{-s\eta^\star(z_j, \ub)} \rho(ds \mid z_j) = e^{-s\{1 + \eta^\star(z_j, \ub)\}}, \, j = 1, \dots, n$. The discrete nature of $\tmu$ introduces ties in $z_i$. Let $\{z^\star_1, \dots, z^\star_k\}$ denote the unique values, with multiplicities $\{n^\star_1, \dots, n^\star_k\}$, among the currently imputed $\{z_1,\ldots,z_n\}$. 
For fixed $\bth$, we thus have
$$
P_{J_\ell}\left(s \mid \ub, \bth, z^\star_\ell, \ns_\ell\right) \propto s^{\ns_\ell - 1}\, e^{- s\left\{1 + \sum_{i=1}^n u_i \exp(\th_i z^\star_\ell)\right\}} \equiv \Ga\left(\ns_\ell, \, \psi(z^\star_\ell; \ub, \bth) + 1\right), \ \ell = 1, \dots, k,
$$ with $\psi(z^\star_\ell; \ub, \bth) = \sum_{i=1}^n u_ie^{\th_i z^\star_\ell}$. This completes the proof.
\begin{remark}
\label{remark1}
Eq. (\ref{sup:eq4}) expresses the Laplace functional of $[\mu \mid \zb]$ (with fixed $\bth$) as
$$
\E_\tmu\left(e^{-\int_\sy h(z) \tmu(dz)} \mid \zb\right) = \int_{{\R^{+}}^n} \E_\tmu\left\{e^{-\int_\sy h(z) \tmu (dz)} \mid \zb, \ub \right\} \,   p(\ub \mid \zb) d\ub,
$$
i.e., a mixture of the conditional Laplace functionals of $[\mu\mid\zb,\ub, \bth]$ with mixing measure $p_{\bth}(\ub\mid\zb) \propto D(\ub) e^{-\psi(\ub)}$.
This representation implies that the proposal for $\mu$ (with fixed $\bth$) can be generated via the following two-step procedure:
\begin{enumerate}
    \item Sample $\ub$ from $p_{\bth}(\ub \mid \zb)$.
    \item Conditional on the sampled $\ub$ and for fixed $\bth$, generate a proposal $\tmu$ according to the conditional distribution of $[\mu \mid \zb, \ub, \bth]$, as specified by its Laplace transform.
\end{enumerate}
Subsequently, the proposed $\tmu$ is accepted or rejected based on MH acceptance probability computed using the marginal (w.r.t. $\ub$) proposal density $q_{\bth}(\tmu \mid \zb)$ (for details on $q$, see \emph{$\tmu$ update} in Appendix \ref{sec:mcmc}).  
\end{remark}

\paragraph*{Proof of Proposition \ref{result2}.}
From Eq. (\ref{sup:eq4}) and Remark \ref{remark1}, for fixed $\bth$ the posterior for $\ub = (u_1, \dots, u_n)$ conditional on $\zb$ is given by
\begin{align}
\label{sup:eq5}
p_{\bth}(\ub \mid \zb) \propto D(\ub) e^{-\psi(\ub)}\ ,
\end{align}
where $D(\ub) = \prod_{j=1}^n \int_{\R^{+}} s e^{-s \eta^\star(z_j, \ub)} \rho(ds \mid z_j)$ and 
$e^{-\psi(\ub)} = e^{\left[ - \int_{S}  \left\{1 - e^{- \eta^\star(z, \ub) s} \right\} \nu(ds, dz) \right]}$. By standard Laplace transform theory, we get
\begin{align*}
e^{-\psi(\ub)} & =  \exp \left[ - \int_{\R^{+} \times \sy}   \left\{1 - e^{- s\sum_{i=1}^n u_i \exp(\th_i  z)} \right\}\frac{e^{-s}}{s}ds \, \alpha  G_0(dz) \right]\\
& =  \exp \left[ - \alpha \int_{\sy} \ln \left\{1+ \sum_{i=1}^n u_i \exp(\th_i  z) \right\} G_0(dz) \right]
\end{align*}
Similarly, one may show that 
$$
D(\ub) =  \prod_{j=1}^n \left\{1+ \sum_{i=1}^n u_i \exp(\th_i  z_j)\right\}^{-1} = \exp\left[- \sum_{j=1}^n \ln \left\{ 1+ \sum_{i=1}^n u_i \exp(\th_i  z_j)\right\}  \right].
$$ 
Thus, combining these results we obtain, for fixed $\bth$,
\begin{align}
\label{sup:eq6}
p(\ub \mid \bth, \zb) & \propto \exp \left[- \int_{\sy} \ln \left\{1+ \sum_{i=1}^n u_i \exp(\th_i  v) \right\} G_n(dv) \right], 
\end{align}
where $G_n = \alpha G_0 +  \sum_{j=1}^n \delta_{z_j}$. The discrete nature of $\tmu$ introduces ties in $z_i$. Let $\{z^\star_1, \dots, z^\star_k\}$ denote the unique values, with multiplicities $\{n^\star_1, \dots, n^\star_k\}$, among the currently imputed $\{z_1,\ldots,z_n\}$. Clearly, $\sum_{\ell=1}^k n^\star_\ell = n$. Then $G_n$ in (\ref{sup:eq6}) can as well be written as $G_n = \alpha G_0 + \sum_{\ell = 1}^k n^\star_\ell \delta_{z^\star_\ell}$.  This completes the proof.

\section{Posterior MCMC}
\label{sec:mcmc}
We specify the transition probabilities for the Markov chain Monte Carlo (MCMC) posterior simulation under the proposed DPGLM model, using the notations introduced in (\ref{DPGLM}). The model parameters $(\beta, \tmu)$, the latent variables $\zb = (z_1, \dots, z_n), \ub = (u_1, \dots, u_n)$, and the derived parameters $\bth = (\th_1, \dots, \th_n)$ represent the currently imputed values. Here, $y_i, z_i \in \sy$ and the mean function is given by
$\lambda_i = \lambda(x_i) =  \E(z_i \mid x_i) = g^{-1}(\eta_{x_i})$,
where $\eta_x$, for example, can be a linear predictor $x^T\beta$. The derived parameter is obtained via $\th_x = {b^\prime}^{-1}\{\lambda(x)\}$. Let $H$ denote the finite truncation point in the Ferguson--Klass algorithm for approximating the CRM. In the discussion that follows, we use the notation ``$\ldots$" in the conditioning set of complete conditional posterior distributions to indicate all other (currently imputed) parameters and the data. 

\paragraph*{Step 1: $\beta$ update.} The complete conditional for $\beta$ is $\pi(\beta \mid \zb, \tmu,\rest)$, with 
\begin{align*}
    \log \pi(\beta \mid \zb, \tmu,\rest) = \sumni \{\th_i z_i - b(\th_i,\tmu) + \log \tmu(z_i)\} + \log p(\beta),
\end{align*}
where $p(\beta) \equiv \Normal(\mu_\beta, \Sigma_\beta)$ is the
$\beta$ prior and $b(\th_i, \tmu) = \log \int \exp(\th_i
v) \tmu(dv)$ is the log-normalizing constant. We update $\beta$ by first obtaining the posterior
mode,  $\beta^\star = \arg \max_{\beta} \log \pi(\beta \mid \zb, \tmu,
\rest)$,  
and using the proposal $\widetilde \beta \sim \Normal(\beta^\star,
\Sigma^\star)1_A\left(\widetilde{\beta}\right)$, where $\Sigma^\star =
\left\{\sum_{i=1}^n  \frac{x_i
    x_i^T}{\left(g^{\prime}\left(\lambda_i\right)\right)^2 b^{\prime
      \prime}\left(\th_i\right)}\right\}^{-1}$ is the inverse Fisher
information at $(\beta^\star, \tmu)$ and $A = \left\{\beta \in \R^p :
  \lambda_i \in \sy, \frall i \right \}$. The proposal is then
accepted or rejected via a Metropolis-Hastings (MH) step.

\paragraph*{Step 2: $\ub$ update.} For fixed $\bth$, the posterior for $\ub = (u_1, \dots, u_n)$ conditional on $\zb$ is 
\begin{eqnarray*}
  p(\ub \mid \zb, \bth) \propto \exp \left \{ - \int_{\sy} \log \left[1+
  \sum_{i=1}^n u_i \exp(\th_i  v) \right] G_n(dv) \right\},  
\end{eqnarray*}
where $G_n = \alpha G_0 +  \sum_{j=1}^n \delta_{z_j}$. Let $\{z^\star_1, \dots, z^\star_k\}$ denote the unique values, with multiplicities $\{n^\star_1, \dots, n^\star_k\}$, among the currently imputed $\{z_1,\ldots,z_n\}$. Then $G_n$ can be written as $G_n =
\alpha G_0 + \sum_{\ell = 1}^k n^\star_\ell \delta_{z^\star_\ell}$.
Following \citet{barrios&al:13}, we generate a proposal using random
walk, $\widetilde u_j \sim \text{ Gamma}\left(\delta,
  \delta\big/u_j\right)$ and follow up with an MH
acceptance step. The tuning parameter \ech $\delta (\geq 1)$
controls the acceptance rate of the MH step.

\paragraph*{Step 3: $\tmu$ update.} Conditional on $\ub$ and for fixed $\bth$ the posterior on $\tmu$ is a inhomogeneous CRM, as described in Proposition 
\ref{result3}.
However, $\bth$ depends on $\tmu$. We therefore can not use the result for a Gibbs sampling transition probability (updating $\tmu$ by a draw from the complete conditional posterior). Instead we use Proposition~\ref{result3} to implement an MH transition probability. 
For the following discussion, let $\pi(\tmu)$ denote the target posterior distribution of $\tmu$.
 We assume that $\muc$ in Proposition \ref{result3} is generated using the Ferguson-Klaas
algorithm truncated at a fixed number of $H$ atoms selected by decreasing weights. In that case, the proposal distribution $q$ as well as the target posterior distribution $\pi(\tmu)$ reduce to finite-dimensional distributions with
density w.r.t.  Lebesgue measure, allowing us to construct an MH transition probability.
Let then $q_{\bth(\tmu)}(\tmus \mid \ub, \zb)$ denote the inhomogeneous CRM described in Proposition \ref{result3}. We generate a proposal $\tmus \sim q$.
In the following expression we will need normalization constants with
different combinations of the CRM $\tmu$ and exponential tilting based
on arbitrary $\th_x$, with $\th_x \ne \th_x(\tmu)$, that is,
different from the derived parameter under $\tmu$.
We therefore introduce notation   
$$
b(\th_x,\tmu)=  \log \int_{\YY} \exp (\th_x v) \tmu(dv)
$$ 
to denote the log normalization constant when CRM $\tmu$ is used with
exponential tilting based on $\th_x$. 
The proposal $\tmus \sim q$ is then followed up with a Metropolis-Hastings acceptance/rejection step with acceptance ratio
$$
r = \prod_i \exp\{2(\th^
\star_i - \th_i) z_i - b(\th^\star_i, \tmu^\star) + b(\th_i, \tmu) - b(\th^\star_i, \tmu) + b(\th_i, \tmu^\star)\},
$$
where $\th^\star_i = \th_{x_i}(\beta, \tmu^\star)$ and accept the
proposal with probability $r \wedge 1$ (see Section \ref{A21} for a derivation of $r$). 
The details of generating a proposal $\tmus \sim q$ are as follows.
In Proposition~\ref{result3}, for fixed $\bth = \bth(\tmu)$,
we generate 
$\tmus \stackrel{d}{=} \muc + \sum_{\ell=1}^k J_\ell
\delta_{z^\star_\ell} \sim q_{\bth}(\tmus \mid \ub, \zb)$, where  
\begin{itemize}
 \item[(a)]  $\muc \sim \text{ CRM}\left(\nuc\right) \text{ with the L\'evy intensity }
   \nuc(ds, dz) =  \frac{e^{- s\left(\psi(z) +1\right)}}{s} ds \ 
   \alpha  G_0(dz)$, where $\psi(z) = \sum_{i=1}^n u_i \exp(\th_i z)$.
   We write $\muc = \sum_{h =1}^H s_h \delta_{\bar z_h}$. The
   random atom locations $\bar z_h$ and weights $s_h$
   are generated using the \citet{ferguson1972representation}
   algorithm: it first generates the random weights $s_h$ in decreasing order. For that, we sample $\xi_h \sim$ standard Poisson process (PP) of unit rate
   i.e. $\xi_1, \xi_2 - \xi_1, \dots \iid \text{ Exp}(1)$. Then solve for
   $s_h = N^{-1}(\xi_h)$, with  
 $$
 N(v) = \nuc([v, \infty], \sy) = \int_v^\infty  \int_{\sy} \nuc(ds, dz) = \alpha \int_v^\infty  \int_{\sy} \frac{e^{- s\left(\psi(z) +1\right)}}{s} G_0(dz)\ ds
 $$
 Next, the random atom locations $\bar z_h$ are sampled from the conditional cumulative distribution function $F_{\bar z_h \mid s_h = s}(z) = \frac{\nuc\left(ds, \ (-\infty, z]\right)}{\nuc(ds, \, \sy)}$, which is conditional on the generated random weights $s_h$. 
 \item[(b)] Let $\{z^\star_1, \dots, z^\star_k\}$ denote the unique values, with multiplicities $\{n^\star_1, \dots, n^\star_k\}$, among the currently imputed $\{z_1,\ldots,z_n\}$. Then for fixed
   atom $z^\star_\ell$, the random weight $J_\ell$ is
   generated from $P_{J_\ell}(s \mid \rest)\propto s^{n^\star_\ell - 1}
   e^{- s\{\psi(z^\star_\ell) +1\}} \equiv \text{ Gamma}
   \left(n^\star_\ell,  \psi(z^\star_\ell) +1\right), \ \ell = 1,
   \dots, k$, where $\psi(z^\star_\ell) = \sum_{i=1}^n u_i \exp(\th_i z^\star_\ell)$. 
\end{itemize}
\paragraph*{Step 4: $\zb$ update.} The complete conditional for $z_i$ is
\begin{eqnarray*}
\pi(z_i \mid \rest) \propto K(y_i \mid z_i) \sum_{\ell} 
 \exp(\theta_i z_i) {\tilde{J}_\ell} \delta_{{\tilde z}_\ell}(z_i),
\end{eqnarray*} 
where $\mathcal{Z} := \{{\tilde z}_\ell\}_{\ell \geq 1} = \{\bar z_1, \dots, \bar z_H, z^\star_1, \dots,  z^\star_k\}$ and $\mathcal{J} := \{{\tilde{J}}_\ell \}_{\ell \geq 1} = \{s_1, \dots, s_H, J_1, \dots,  {J_k}\}$.

\subsection{The MH acceptance ratio $r$}
\label{A21}
Let $\zb, \beta, \tmu$ be the currently imputed values, with derived parameters $\bth = \bth(\beta, \tmu)$, and log-normalizing constants $b(\th_i, \tmu) = \log \int \exp(\th_i v) \tmu(dv)$. The target posterior for $\tmu$ is $\pi(\tmu \mid\beta, \zb, \rest) = \left\{\prod_i e^{\th_i z_i - b(\th_i, \tmu)} \tmu(z_i)\right\} \, p(\tmu),$ which, for notational simplicity, we denote by $\pi(\tmu)$.

Let $\tmus$ denote the proposal, with density $q_{\bth(\tmu)}(\tmu^* \mid \zb) = \left\{\prod_i e^{\th_i z_i - b(\th_i, \, \tmu^\star)} \, \tmu^*(z_i)\right\}\, p(\tmu^*),$ which we denote as $q(\tmus \mid \tmu)$ for simplicity. The derived parameters corresponding to $\tmus$ are given by $\bth^* = \bth(\beta, \tmu^*)$. 
Note again that in the proposal density the factor $\th_i$ in the exponential tilting is {\em not} matching the derived parameter $\th^\star_i$. The Metropolis--Hastings acceptance ratio is given by $r = \frac{\pi(\tmu^*)}{\pi(\tmu)} \frac{q(\tmu \mid \tmu^*)}{q(\tmu^* \mid \tmu)}$. Substituting the expressions for $\pi$ and $q$, we have
\begin{align*}
    r  &= \frac{\prod_i e^{\th^*_i z_i - b(\th^*_i, \, \tmu^*)} \tmu^*(z_i) \, p(\tmu^*)}{\prod_i e^{\th_i z_i - b(\th_i, \, \tmu)} \tmu(z_i) \, p(\tmu)} \frac{\prod_i e^{\th^*_i z_i - b(\th^*_i, \, \tmu)} \tmu(z_i) \, p(\tmu)}{\prod_i e^{\th_i z_i - b(\th_i, \, \tmu^*)} \tmu^*(z_i) \, p(\tmu^*)} \\
    &= \prod_i \exp\{2(\th^*_i - \th_i) z_i - b(\th^*_i, \, \tmu^*) + b(\th_i, \, \tmu) - b(\th^*_i, \, \tmu) + b(\th_i, \, \tmu^*)\}.
\end{align*}

\newpage
\section{Additional Simulation Results}
\label{sec:additional_simulation}

\begin{figure}[!ht]
    \centering
    \includegraphics[width=0.8\linewidth, height = 13cm]{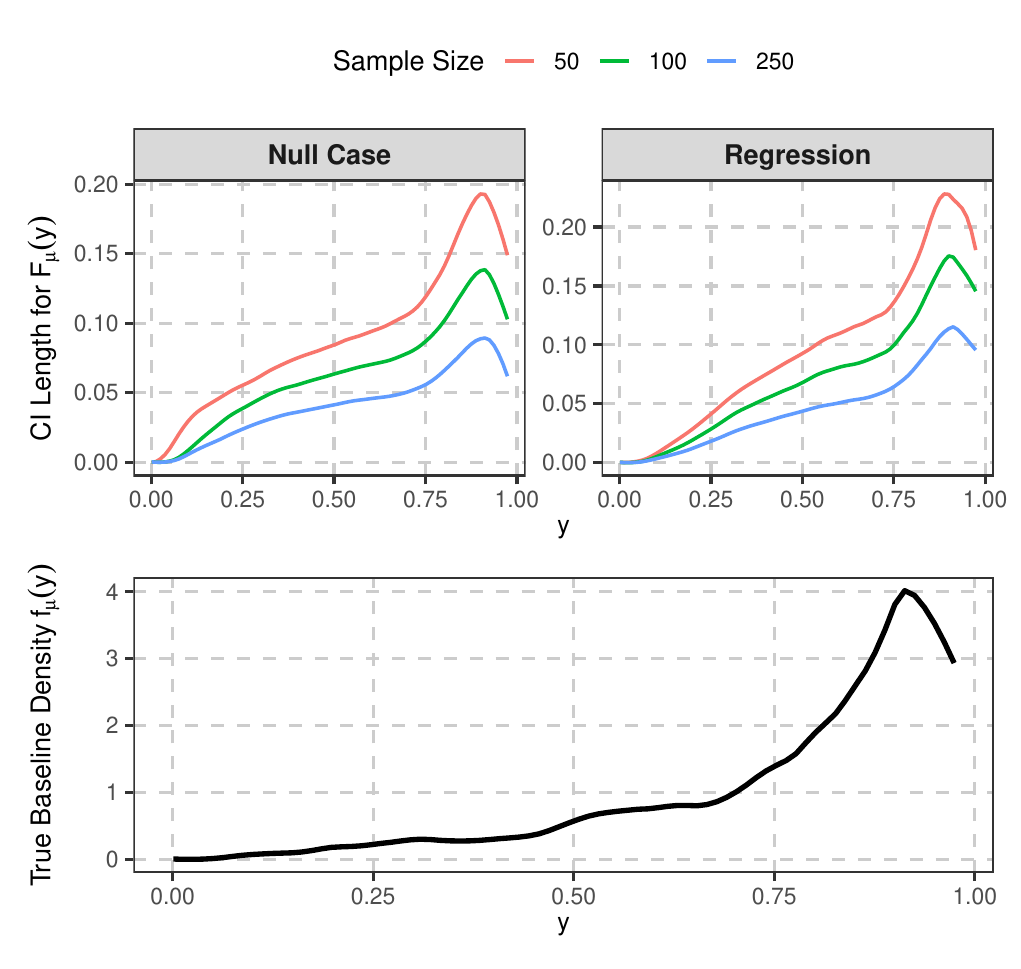}
    \caption{Pointwise credible interval lengths for the baseline CDF $F_{\tmu}(y)$, on a grid of $y$ values, for varying sample sizes, across both simulation scenarios: \emph{Null Case} (upper left) and \emph{Regression} (upper right). The bottom panel displays the true baseline density $f_{\tmu}$. Results are based on \XX simulated data replicates.}
    \label{fig:supp1}
\end{figure}

Figure~\ref{fig:supp1} shows the average pointwise credible interval lengths for $F_{\tmu}(y)$, computed on a grid of $y$ values under both simulation scenarios. The narrowing of these intervals with increasing sample size indicates improved precision.

\begin{table}[!ht]
    \centering
    \caption{Bias, Root-mean-square error (RMSE), coverage probability (\%), and credible interval (CI) length for the \emph{exceedance probability, $p(y > y_0 \mid x)$} at $x = (1, 0)$, evaluated at various quantiles $y_0$. Results are based on second simulation scenario and \XX simulated data replicates.}
    \label{tab:exc_reg_1}
    \vspace{0.3cm}
    \begin{tabular}{l c c c c c c}
        \toprule
        \multirow{2}{*}{Metric} & \multirow{2}{*}{Sample Size ($n$)} & \multicolumn{5}{c}{Quantiles ($y_0$)}\\ 
        \cline{3-7} 
        &  & 10\% & 25\% & 50\% & 75\% & 90\% \\
        \midrule
        \multirow{3}{*}{Bias} 
            & 50  & -0.01 & 0.00 & 0.00 & 0.00 & 0.00 \\
            & 100 & 0.00  & 0.00 & 0.01 & 0.00 & 0.00 \\
            & 250 & 0.00  & 0.00 & 0.00 & 0.00 & 0.00 \\
        \midrule
        \multirow{3}{*}{RMSE} 
            & 50  & 0.04  & 0.06 & 0.08 & 0.07 & 0.04 \\
            & 100 & 0.03  & 0.04 & 0.05 & 0.05 & 0.03 \\
            & 250 & 0.02  & 0.03 & 0.03 & 0.03 & 0.02 \\
        \midrule
        \multirow{3}{*}{Coverage (\%)} 
            & 50  & 96.5  & 92.5 & 91   & 91   & 90.5 \\
            & 100 & 92.5  & 92   & 89   & 90   & 94 \\
            & 250 & 90    & 91   & 93   & 93   & 92 \\
        \midrule
        \multirow{3}{*}{CI Length} 
            & 50  & 0.15  & 0.22 & 0.26 & 0.22 & 0.14 \\
            & 100 & 0.10  & 0.16 & 0.18 & 0.16 & 0.10 \\
            & 250 & 0.07  & 0.10 & 0.12 & 0.10 & 0.06 \\
        \bottomrule
    \end{tabular}
\end{table}

\begin{table}[!ht]
    \centering
    \caption{Bias, Root-mean-square error (RMSE), coverage probability (\%), and credible interval (CI) length for the \emph{exceedance probability, $p(y > y_0 \mid x)$} at $x = (1, 0.25)$, evaluated at various quantiles $y_0$. Results are based on second simulation scenario and \XX simulated data replicates.}
    \label{tab:exc_reg_2}
    \vspace{0.3cm}
    \begin{tabular}{l c c c c c c}
        \toprule
        \multirow{2}{*}{Metric} & \multirow{2}{*}{Sample Size ($n$)} & \multicolumn{5}{c}{Quantiles ($y_0$)}\\ 
        \cline{3-7} 
        &  & 10\% & 25\% & 50\% & 75\% & 90\% \\
        \midrule
        \multirow{3}{*}{Bias} 
            & 50  & -0.01 & 0.00 & 0.00 & 0.00 & 0.00 \\
            & 100 & 0.00  & 0.01 & 0.00 & 0.00 & 0.00 \\
            & 250 & 0.00  & 0.00 & 0.00 & 0.00 & 0.00 \\
        \midrule
        \multirow{3}{*}{RMSE} 
            & 50  & 0.04  & 0.07 & 0.08 & 0.07 & 0.04 \\
            & 100 & 0.03  & 0.05 & 0.06 & 0.05 & 0.03 \\
            & 250 & 0.02  & 0.03 & 0.03 & 0.03 & 0.02 \\
        \midrule
        \multirow{3}{*}{Coverage (\%)} 
            & 50  & 95.5  & 93.5 & 87   & 92   & 87 \\
            & 100 & 93    & 91   & 90   & 90   & 93 \\
            & 250 & 93    & 93   & 94   & 92.5 & 92 \\
        \midrule
        \multirow{3}{*}{CI Length} 
            & 50  & 0.15  & 0.23 & 0.26 & 0.23 & 0.14 \\
            & 100 & 0.10  & 0.16 & 0.19 & 0.16 & 0.10 \\
            & 250 & 0.06  & 0.10 & 0.12 & 0.10 & 0.07 \\
        \bottomrule
    \end{tabular}
\end{table}

\end{document}